\documentclass[letterpaper]{article}

\usepackage{natbib,alifeconf}  

\usepackage{booktabs}
\usepackage{graphicx} 
\usepackage{mathtools}
\usepackage{amsfonts}
\usepackage{booktabs}
\usepackage{url,hyperref,cleveref}
\usepackage{xcolor}
\usepackage{soul}
\usepackage{tikz}
\usepackage{multirow}

\newcommand{\celeg}{{\it C. elegans}}
\title{ Behavioural Classification in \celeg: a Spatio-Temporal Analysis of Locomotion}
\author{
    Nemanja Antonic$^{1}$,
    Monika Scholz$^{2}$,
    Aymeric Vellinger$^{1}$,
    Euphrasie Ramahefarivo$^{2}$, \and
    Elio Tuci$^{1}$ \\
    \mbox{}\\
    $^1$University of Namur, Belgium \\
    $^2$Max Planck Institute for Neurobiology of Behaviour, Germany\\
    nemanja.antonic@unamur.be
} 
\begin{document}

\maketitle
\begin{abstract}
The $1$mm roundworm \celeg\ is a model organism used in many sub-areas of biology to investigate different types of biological processes. In order to complement the {\it in-vivo} analysis with computer-based investigations, several methods have been proposed to simulate the worm behaviour. These methods extract discrete behavioural units from the flow of the worm movements using different types of tracking techniques. Nevertheless, these techniques require a clear view of the entire worm body, which is not always achievable. For example, this happens in high density worm conditions, which are particularly informative to understand the influence of the social context on the single worm behaviour. In this paper, we illustrate and evaluate a method to extract behavioural units from recordings of \celeg\ movements which do not necessarily require a clear view of the entire worm body. Moreover, the behavioural units are defined by an unsupervised automatic pipeline which frees the process from predefined assumptions that inevitably bias the behavioural analysis. The behavioural units resulting from the automatic method are interpreted by comparing them with hand-designed behavioural units. The effectiveness of the automatic method is evaluated by measuring the extent to which the movement of a simulated worm, with an agent-based model, matches the movement of a natural worm.
Our results indicate that spatio-temporal locomotory patterns emerge even from single point worm tracking. Moreover, we show that such patterns represent a fundamental aspect of the behavioural classification process.
\end{abstract}



\section{Introduction}
Defining and modelling atomic behaviours in the flow of movements of an animal requires filtering the intrinsic complexities of life, by arbitrarily segmenting a continuous spatio-temporal process into discrete units of behaviour. In the roundworm \celeg, this process has been attracting the interest of the research community since a clear characterisation of the worm behavioural repertoire may help identify underlying neurophysiological mechanisms within its relatively small ($302$ neurons in the hermaphrodite condition) and fully mapped (see the connectome) nervous system~\citep[][]{white1986structure}. Different methodological approaches have been already described in the literature to segment worm changes in body postures. The work described in~\citep{openworm2014, openworm2018} tried to decompose the worm movements in a bottom-up fashion, starting from how neurophysiological circuits stimulate muscles and generate locomotion. The resulting model is expectedly computationally expensive, given the detailed nature of the system description, and generates only partially correct \textit{in-silico} models.  A series of other studies~\citep[see][]{Salvador2014, Costa2024MarkovWorm, Broekmans2016} take a different approach by defining behavioural units using a dimensionality reduction framework originally proposed by~\citet{stephens2008}.
These authors first overlay a series of uniformly distributed points along the worm centre line. Subsequently, they connected these points with segments and recorded the variations through time of the angles between these segments. The resulting covariance matrix, representing the worm postural changes, was analysed with the Principal Component Analysis. Their results showed that just $4$ dimensions (i.e., eigenvalues, later termed \textit{eigenworms}) account for $95\%$ of the worm's posture changes. The eigenworm analysis described above comes with several drawbacks. First of all, self-occluding shapes make it difficult to clearly delimit the worm centre line, required to build the covariance matrix~\citep{Broekmans2016}. In order to overcome this limitation, alternative methods rely on tracking the worm centre-of-mass, for example by labelling parts of the worm body with fluorescent markers~\citep{bonnard2022}. The drawback of relying on centre-of-mass to define behaviour is that key navigational elements such as omega (i.e., when a worm touches its tail with its head, resulting in an $\Omega$ shape) and delta turns (i.e., a partial omega turn, resembling a $\delta$ shape) can be hardly discerned.

A second drawback of the eigenworm analysis lies in the acquisition and processing of data in high density worm conditions, in which a clear view of single worm bodies during tracking becomes unfeasible. Nevertheless, these conditions are the most informative to infer whether and eventually how the social context influences single worm behaviour, and which individual responses contribute to the emergence of collective and coordinated actions in large groups of worms. Collective behaviour in populations of \celeg\ is the focus of the EU funded project BABots~\citep{BABOTS_WebPage}, which aims to reprogram the neural circuits of these worms through genetic modifications to make them capable of carrying out collaborative and coordinated collective responses that are not necessarily part of their natural repertoire. In other words, the BABots project develops theories and methods to deal with \celeg\ as if they were {\textbf b}iological {\textbf a}nimal ro{\bf bots} (i.e, babots) to be reprogrammed to execute tasks for our benefit. By working within the BABots project, our long term objective is to propose solid solutions to overcome the different limitations of current methods in the characterisation of \celeg\ behaviour to facilitate the study of neural mechanisms underlying single and collective worm responses.

In this paper, we contribute to this challenge by describing a comparative study which evaluates two methods (i.e., the atomic and the automatic behaviour segmentation) to segment \celeg\ behaviours by processing worm centre of body tracks. In both methods, the flow of worm movements is analysed using kinematic and spatial properties of the trajectory of the worm centre of body. In the atomic method, the behavioural units are defined using hand-designed principles, while in the automatic method these principles emerge from an unsupervised clustering analysis of the data. The effectiveness of these methods is evaluated by simulating the worm movements with the support of a probabilistic finite-state machine in which different states correspond to different behavioural units and the transitions between states are probabilities also extracted from the tracking of the worm.


The automatic method is inspired by the work described in~\citep{Eren2024}, where authors develop an automatic behaviour recognition algorithm, which, from features defined for each data point, is able to produce a set of clustered embeddings, each corresponding to a specific stereotyped behaviour in the nematode \textit{P. pacificus}. A related method is also used in~\citep{Liu2018} for the postural analysis of \celeg\ to quantify the effects of mechanical stimuli.

Approaches similar to the one described in~\citep{Eren2024} exist in the literature to characterise the behaviour of other animals. For example, clustering the posture of freely moving fruit flies reveals a hundred stereotyped behaviours easily mappable to manually defined behaviours~\citep[see][]{berman2014}. In~\citep{Segalin2021}, authors propose a pipeline based on the classification of behaviours from segmented data in mice. More recently, authors in~\citep{Hsu2021, Tillmann2024} have developed fully automated pipelines for the recognition and prediction of single and social animal behaviour from segmented data, mainly for mice and monkeys, combining the automated behaviour extraction with the possibility to include user-defined behaviours. 

The original contributions of our work reside: (i) in showing that the worm trajectories data derived from the relatively simple \celeg' centre of mass analysis is sufficiently rich to characterise the worm behaviour in an effective way; (ii) the automatic behaviour segmentation method outperforms the atomic one.  
In particular, we show that the automatic method is capable of reproducing overall movement more closely to that of a biological worm than the atomic method. Although our results are still preliminary, our primary contribution is that even with the substantial simplification of treating the worm as a moving point, sufficient temporal correlations exist in its movement to determine behavioural states which are effectively identified  by the  automatic method. Compared to previous work, we show that a simpler reading of the tracking leads to an effective extraction of temporal correlations in the worm movement.

The remainder of this article is organised as follows: in Section~\ref{sec:methods}, we define the atomic and the automatic pipelines for defining behaviour at the scale of a point, together with the agent-based model used to assess their performance; in Section~\ref{sec:results} we discuss the comparative results 
in terms of the behavioural units both methods rely on,
and the performance of each behavioural characterisation in the agent-based model; in Section~\ref{sec:conclusion} we draw conclusions and discuss possible future work.

\section{Methods}
\label{sec:methods}


We first define a set of atomic behaviours, together with their defining properties through a hand-designed approach. Then, we describe our pipeline for the automatic recognition of stereotyped behaviours. In particular, we discuss the features of each data point (i.e., a data point corresponds to the worm's centre of mass at a given time step), together with the dimensionality reduction and clustering applied to the feature space (i.e., the space of all features of all data points).  Finally, we describe an agent-based model based on a probabilistic finite-state machine with semi-Markovian dynamics. In this machine, each state is assigned to a behaviour, obtained either through the atomic or the automatic method. The transitions are composed of both a time-dependent and a constant factor, both defined via statistical properties obtained from data.

\subsection{Dataset}
We build a dataset composed of tracking data of $N2$ worms, the most commonly used \celeg\ strain in biological literature. All tracks are from worms moving on food, thus the analysis focuses on worms during feeding. Our dataset is composed of tracks taken from~\citep{Javer2018, Broekmans2016} with a minimum duration of $5$ minutes. 


\subsection{The Atomic Method}
\label{sec:atomic_behaviour}

 A first set of behaviours is derived from the analysis of the movement of a massless point in a $2$D space corresponding to the body centre of \celeg. These behaviours are classified into crawls and reorientarions, as observed in~\citep{Salvador2014}. Crawls are composed of straight or curved runs, referred to as lines and arcs respectively, and circular trajectories, referred to as loops. The reorientations are composed of omega turns, pauses and reversals. Contrary to what discussed by authors in~\citep{Salvador2014}, our set of reorientations does not include pirouettes since pirouettes are defined as a sequence of omega turns and reversals occurring in a relatively short amount of time, therefore they do not represent an atomic behaviour. As far as it concerns omega turns, they require an analysis of the outline of the worm body, which becomes unfeasible in high density scenarios. Therefore, we substitute omega turns with sharp turns, which we define as large ($\ge \frac{2}{3}\pi$) changes in heading. Pauses are defined as points where the speed is lower than a given threshold ($50\mu ms^{-1}$).
 
 We use two methods to identify reversal events during worm locomotion. One method uses the fact that our data is segmented, which means that each point of the worm trajectory contains the coordinates of multiple points of the worm body, allowing us to compare the direction of movement with respect to the direction of the worm head. In particular, we leverage the dot product between the head to body centre and the velocity vectors, which we threshold $(\ge \frac{2}{3}\pi)$. The other method is based on the work in~\citep{hardaker2001}, where the heading of the worm body centre is resampled with respect to the worm body length and thresholded $(\ge \frac{2}{3}\pi)$.


Any point that is not classified as a reorientation is defined as a crawl point. Crawl segments are time-contiguous crawl points. We cluster the logarithm of the crawl segment curvature and angular concordance using the KNN clustering algorithm ($k=3$) to obtain lines, arcs and loops, as in~\citep{Salvador2014}. Let $A$ be the set of crawl segments of a worm. Then, $\forall \mathbf{p} \in A$, the curvature $c_\mathbf{p}$ is defined as the average curvature of the segment $\mathbf{p}$ composed of the vectors $\mathbf{x}, \mathbf{y} \in \mathbb{R}^{|\mathbf{p}|}$:
\begin{equation}
\label{eq:curvature}
    c(\mathbf{p}) = \left<\frac{\left|\mathbf{x}''\mathbf{y}'-\mathbf{x}'\mathbf{y}''\right|}{\left(\mathbf{x}'^2+\mathbf{y}'^2\right)^\frac{3}{2}}\right>
\end{equation}

where the first and second order time derivatives are calculated via a finite difference approximation on subsequent points. Segments where $x_i'^2 + y_i'^2\leq 10^{-3}\mu ms^{-1} | \forall i \in \mathbf{p}$ are assigned $c=-1$ as a sentinel value, so to later assign their logarithm a large value ($10$ times the maximum $c$). The angular concordance $a(\mathbf{p})$ is defined as the square root of the average summed squared cosine and sine of $\theta \in [-\pi, \pi]^{|\mathbf{p}| - 1}$, which is the vector of angles between the consecutive points in $\mathbf{p}$:

\begin{equation}
\label{eq:angular_concordance}
    a(\mathbf{p}) = \sqrt{<\cos^2\theta> +<\sin^2\theta>}
\end{equation}


\subsection{The Automatic Method}
\label{sec:automatic_behaviour}

\begin{table}[b!]
    \centering
    \begin{tabular}{ll}
    \toprule
    \textbf{Parameter} & \textbf{Value} \\
    \midrule
    \texttt{n\_neighbors} & $100$ \\
    \texttt{min\_dist} & $0$ \\
    \texttt{repulsion\_strength} & $10$ \\
    \texttt{negative\_sample\_rate} & $15$ \\
    \texttt{n\_components} & $3$ \\
    \texttt{random\_state} & $42$ \\
    \bottomrule
    \end{tabular}
    \caption{UMAP parameters used to reduce the dimensionality of the feature space to a 3D embedding.}
    \label{tab:umap_params}
\end{table}

The automatic method is based on an approach described in~\citep{Costa2024MarkovWorm, Eren2024} in order to automatically build a set of behavioural states by defining kinematic, spatial and time-dependent features of worm's body points extracted from its trajectories. A core difference from previous work lies in that we focus on the centre point of the worm body, disregarding postural details. For each worm body centre point we look at the following features:
\begin{itemize}
    \item kinematic features: the speed, the angle change, and the interpolated re-sampled angle change with respect to the worm body length;
    \item spatial features: the angular concordance (Equation~\ref{eq:angular_concordance}) and the logarithm of curvature (Equation~\ref{eq:curvature}), both computed on a sliding window of $10$ frames backward and forward in time;
    \item time-dependent features: the lagged speed, the angle change, and the resampled angle change, computed with a lag of $-15$ frames.
\end{itemize}

\begin{figure}
    \centering
    \begin{tabular}{cc}
    \includegraphics[width=0.49\linewidth]{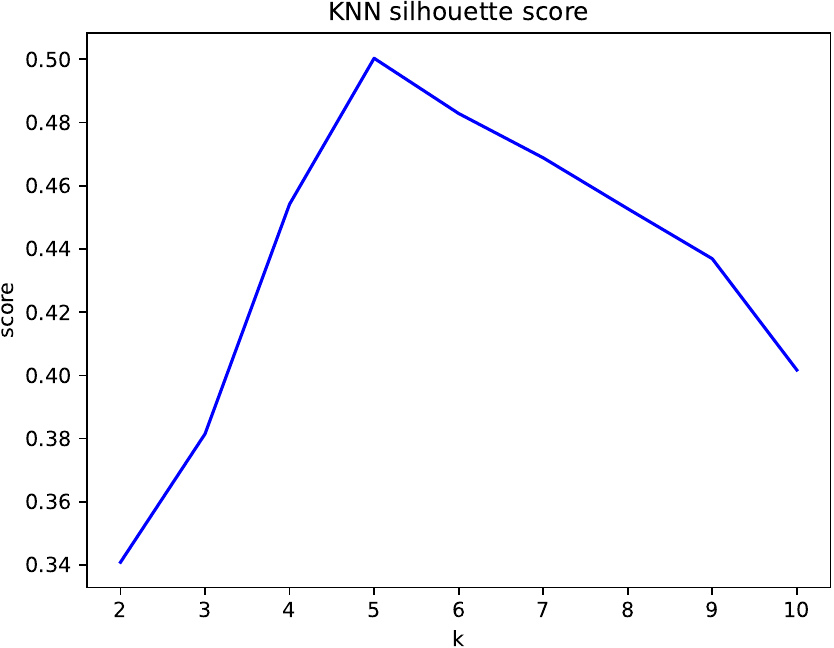}     &  \includegraphics[width=0.49\linewidth]{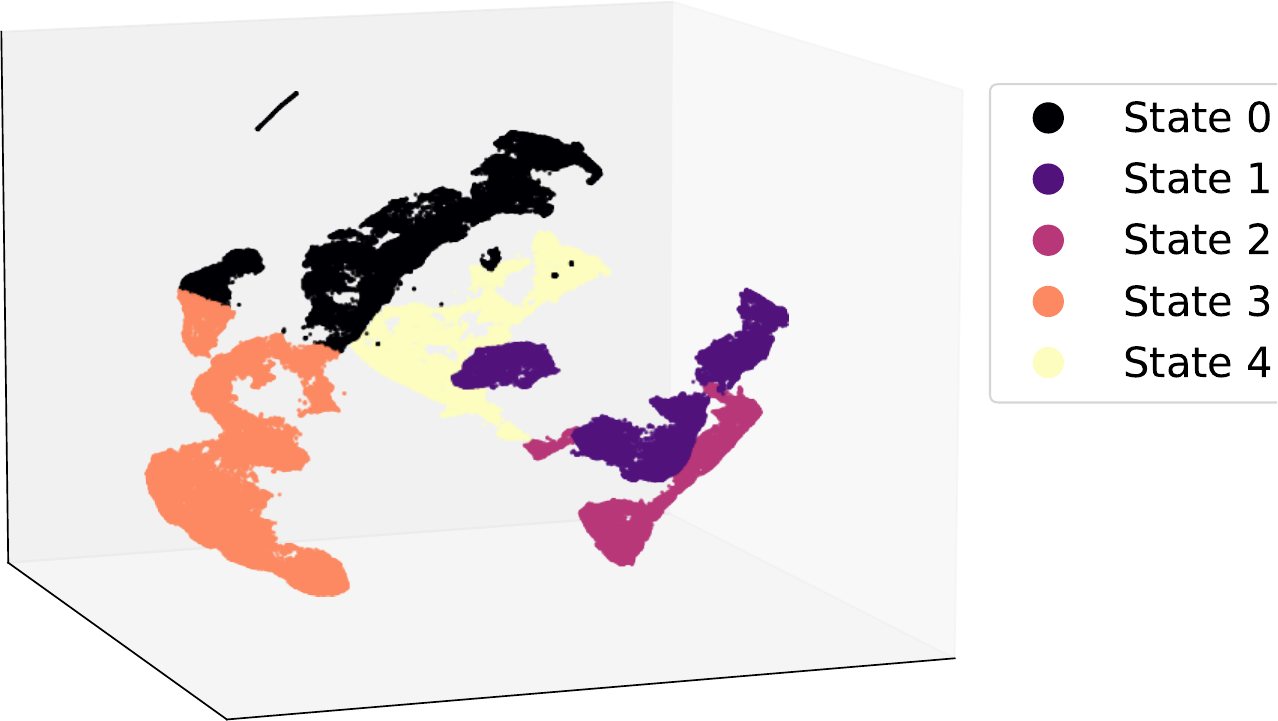}\\
    (a)     &  (b)
    \end{tabular}
 \label{fig:knn_silhouette_embeddings}
 \caption{(a) Average silhouette score obtained from clustering via KMeans the embedded feature space while varying the number of clusters $k$. The maximum value is achieved for $k^*=5$ and the resulting clustering of the embedded feature space is shown in (b). }
\end{figure}

All features are computed by subsampling the data at $4$Hz. Points with a speed exceeding $600\mu ms^{-1}$ are discarded as imaging artifacts. The angle change, the re-sampled angle change, and the lagged angle change whose values are in $[-\pi, \pi]$ are normalised by considering their absolute value, so to take into account only the magnitude of directional change. We proceed by reducing the dimensionality of the normalised feature space by means of UMAP~\citep{McInnes2018} with parameters shown in Table~\ref{tab:umap_params}. Then, we perform a KMeans clustering on the resulting embedded feature space, by selecting the optimal number of clusters $k^* \in [2, 10]$ such that the average silhouette score is maximised (see Figure~\ref{fig:knn_silhouette_embeddings}a). Figure~\ref{fig:knn_silhouette_embeddings}b shows the labelled embedded feature space for the resulting optimal number of clusters $k^*=5$. Each cluster represents a stereotyped behaviour, which we analyse in Section~\ref{sec:results_mapping}.

\subsection{The Agent-Based model}
\label{sec:ab-model}

In our model, each agent is a massless $2$D point with coordinates $x(k),y(k)$ and state $s(k)$ with duration $d(k)$, where $k$ represents the current time step. The worm locomotion is modelled using a finite-state machine in which states represent the behaviours found with the methods described in Sections~\ref{sec:atomic_behaviour} and~\ref{sec:automatic_behaviour}. The finite-state machine is probabilistic since transitions between states are computed by multiplying a constant term by the probability that a given state occurs at each time step $k$. The constant term expresses the average transition rate from one state to another. Each behavioural state is defined by three properties: (i) speed distribution (modelled as a beta distribution); (ii) angle change distribution (modelled as a mixture of Von Mises distributions with varying mean); (iii) duration distribution (modelled as a beta distribution). These distributions are obtained by fitting the speed, the angle change, and the duration of each state from the worm data. During simulation, the agent updates its position by sampling an angle change and a speed from the relative distributions of its current state $s(k)$. Every time step, the duration $d(k)$ of the current state is decreased by one unit until $d(k)=0$. The agent changes to a new state $s(k)$ when the duration of current state expires (i.e., $d(k)=0$).

The transition probability between states is computed as follows:
\begin{equation}
    \label{eq:transition_prob}
        p_{s\rightarrow s'}(k) = (m_{s'}k+q_{s'} ) l_{s\rightarrow s'}
\end{equation}
where the parameters $m_{s'},q_{s'}$ represent a line fit to the mean count of occurrences of state $s'$ binned to windows of $120$ seconds and $l_{s\rightarrow s'}$ is the relative number of transitions from $s$ to $s'$ with respect to the total outgoing transitions from $s$.

\section{Results}
\label{sec:results}
In this Section, we show the results of a series of analyses which tell us whether the two behaviour classification methods described above differ in the characterisation of the worm behaviour, and eventually which one is more effective in modelling the worm movements. Since with the automatic method the worm centre body points are associated to behavioural states in an unsupervised (i.e., automatic) way, 
we first verify 
whether the distributions of speed and heading from the automatic method (see Section~\ref{sec:automatic_behaviour}) match those from the atomic method (see Section~\ref{sec:atomic_behaviour}), 
by using the Shapley values of an XGBoost classifier. The classifier is trained to predict the behavioural state of a data point from its kinematic, spatial and temporal features. The distribution of speed, angle change and duration of each state and the transition probabilities between states define the agent-based model described in Section~\ref{sec:ab-model}. The mean square displacement of agents modelled using the atomic the automatic method are compared between themselves and with the mean square displacement of biological worms from our dataset.

\subsection{Mapping Atomic and Automatic Behaviours}
\label{sec:results_mapping}

\begin{figure*}[t]
    \centering

    \setlength{\tabcolsep}{2pt}
    \renewcommand{\arraystretch}{1.0}
    \begin{tabular}{*{6}{c}} 
        \includegraphics[width=\linewidth/7]{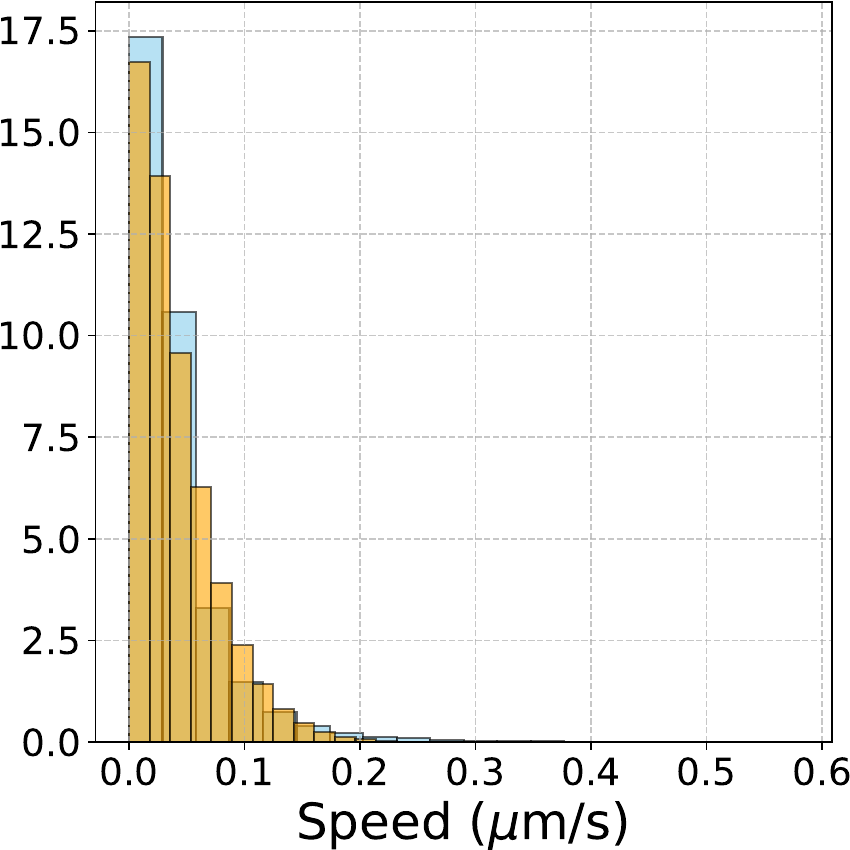} &
        \includegraphics[width=\linewidth/7]{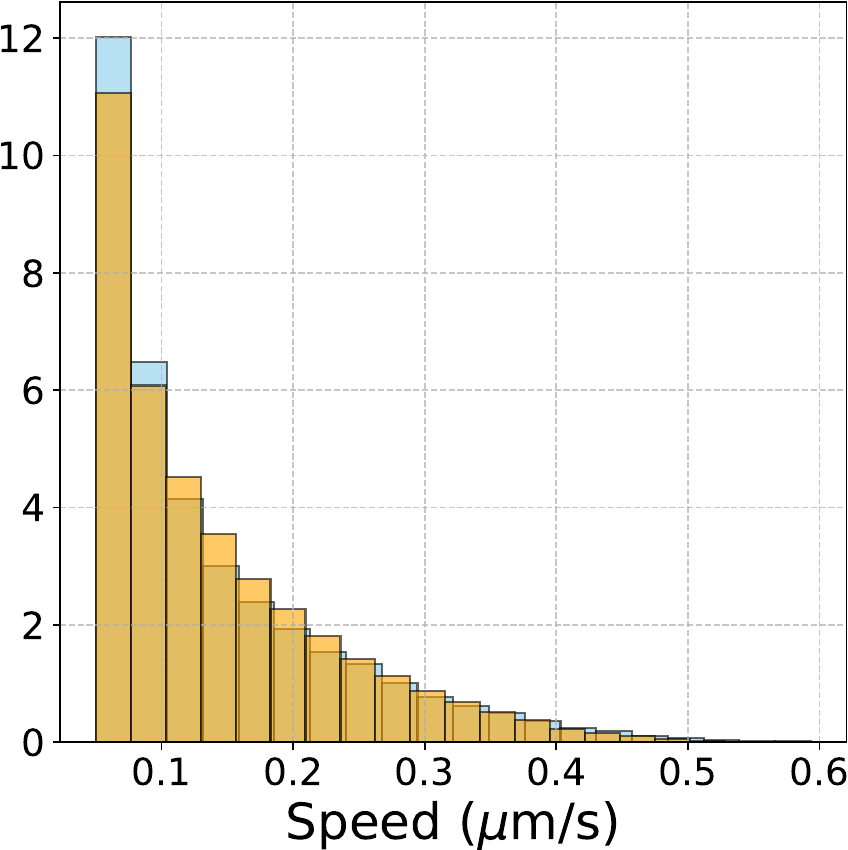} &
        \includegraphics[width=\linewidth/7]{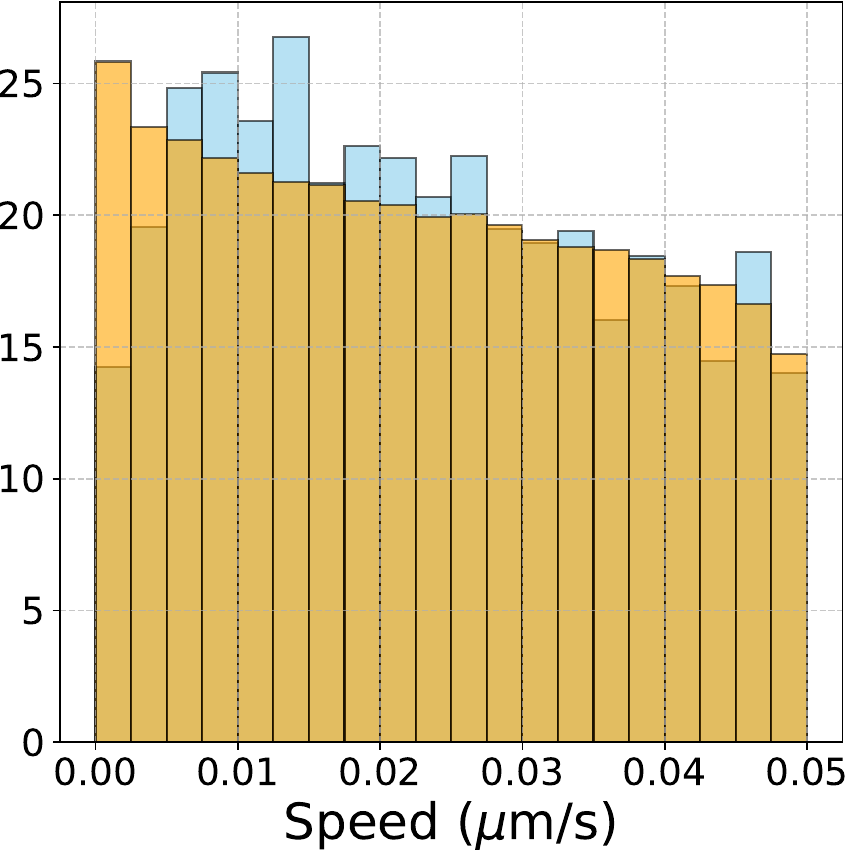} &
        \includegraphics[width=\linewidth/7]{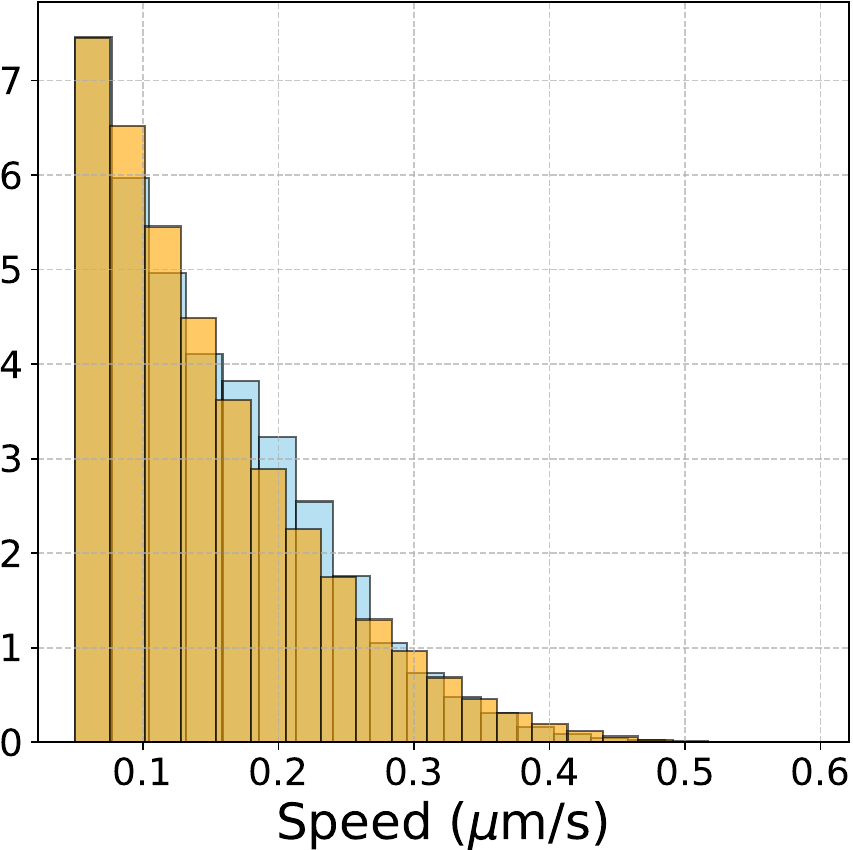} &
        \includegraphics[width=\linewidth/7]{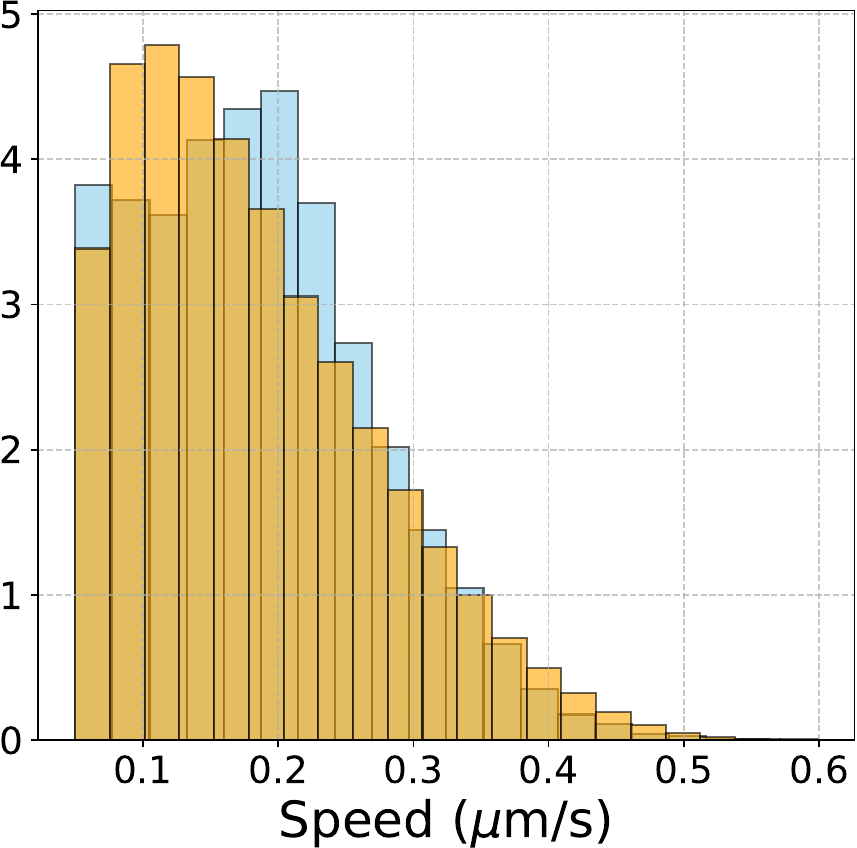} &
        \includegraphics[width=\linewidth/7]{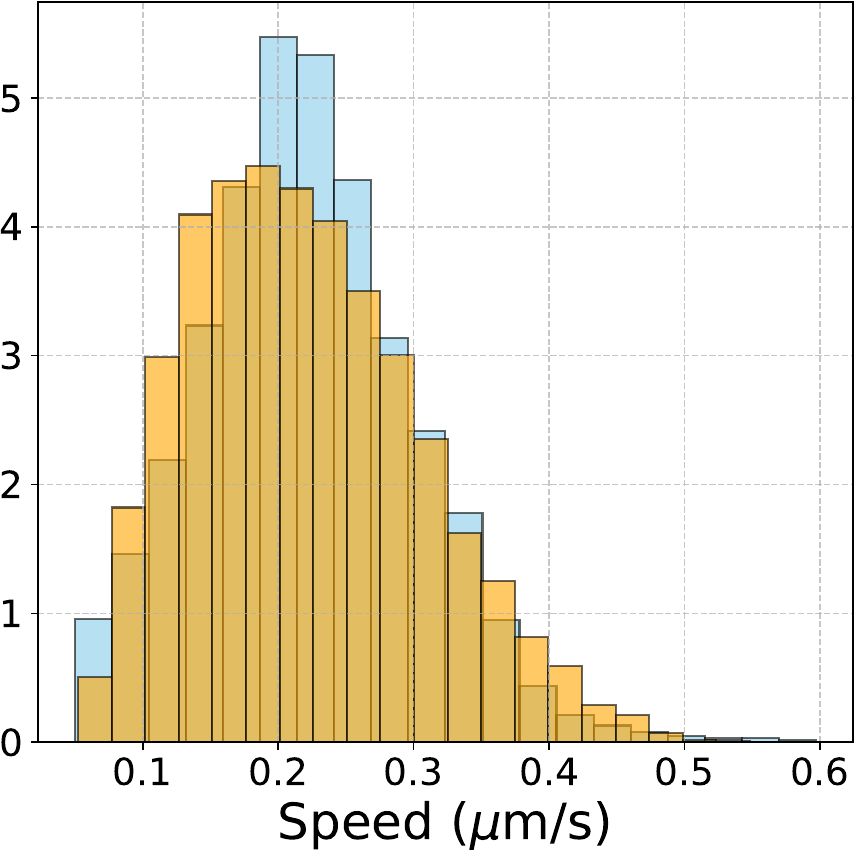} \\
        
        \includegraphics[width=\linewidth/7]{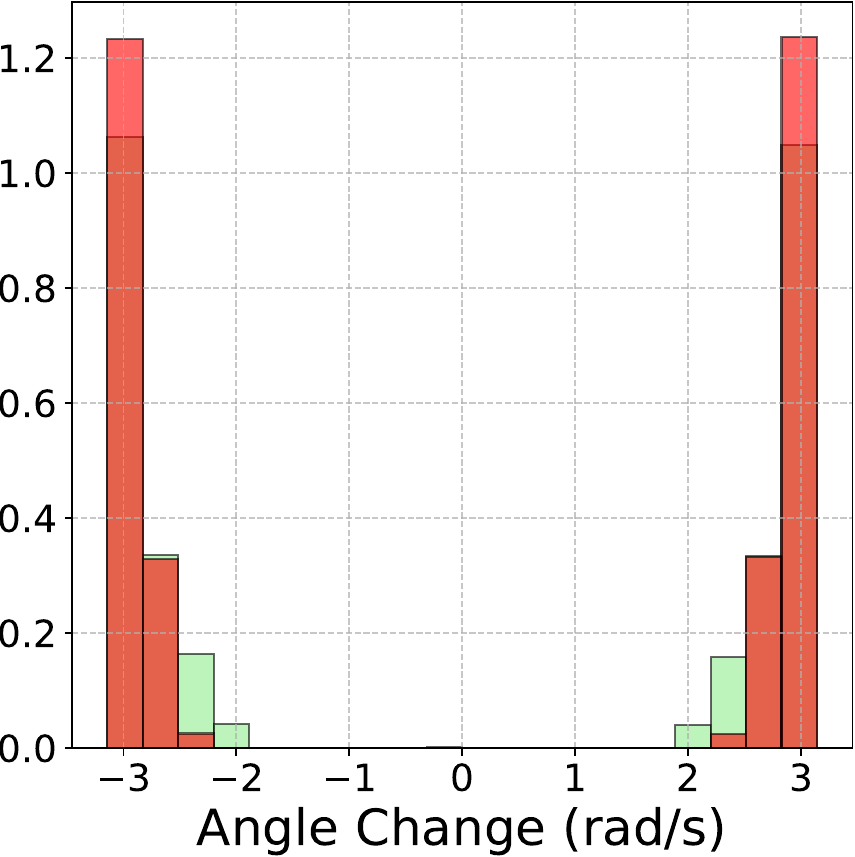} &
        \includegraphics[width=\linewidth/7]{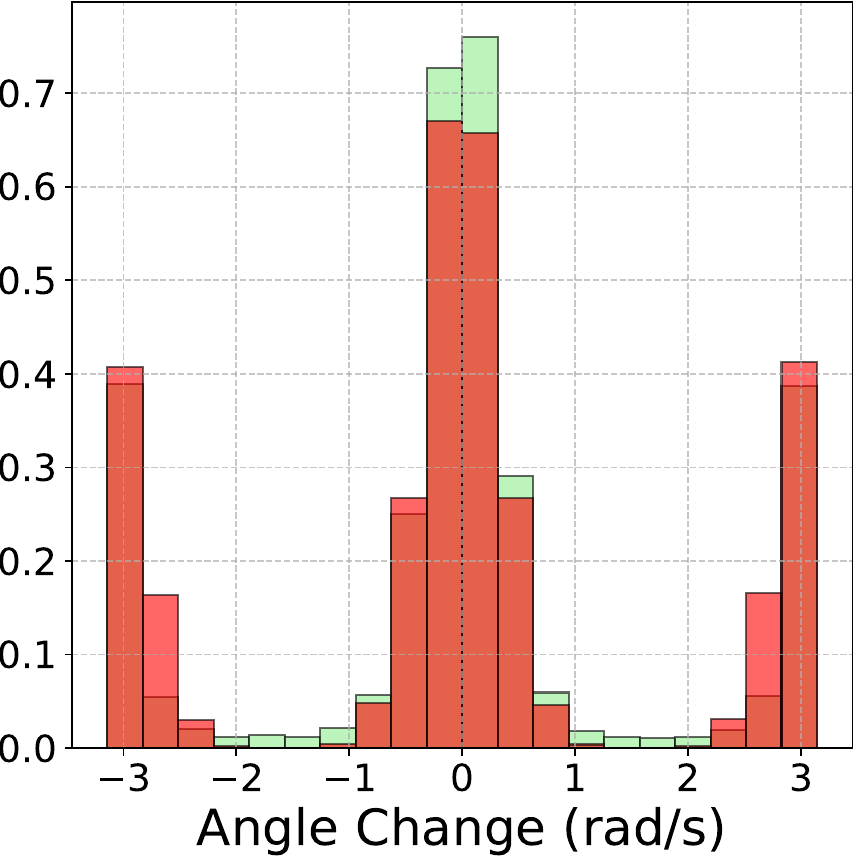} &
        \includegraphics[width=\linewidth/7]{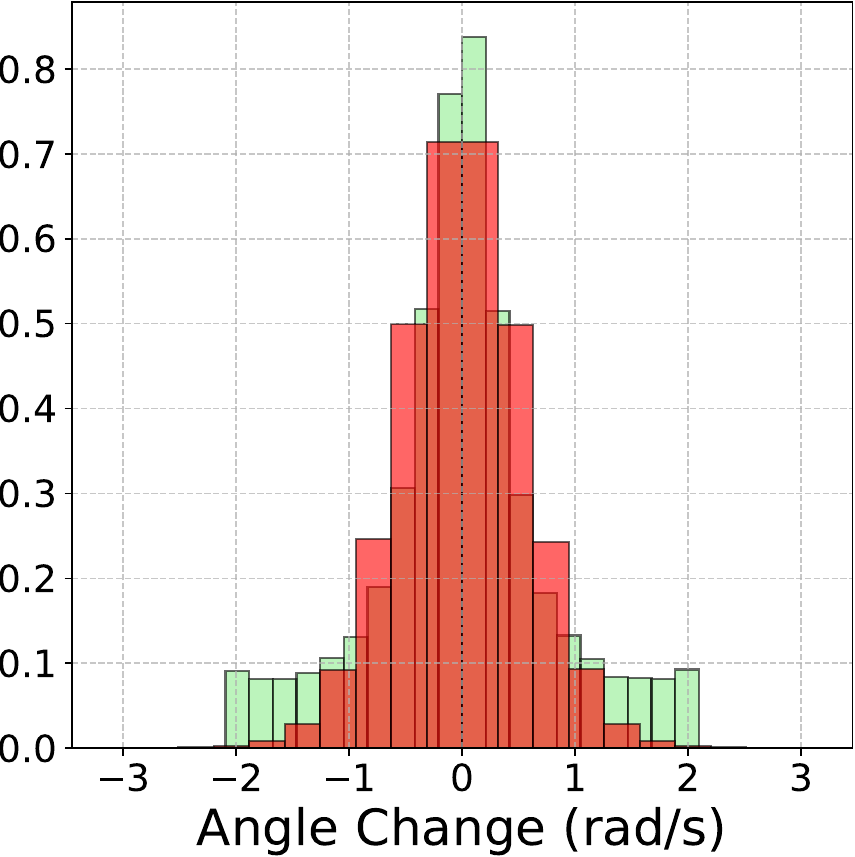} &
        \includegraphics[width=\linewidth/7]{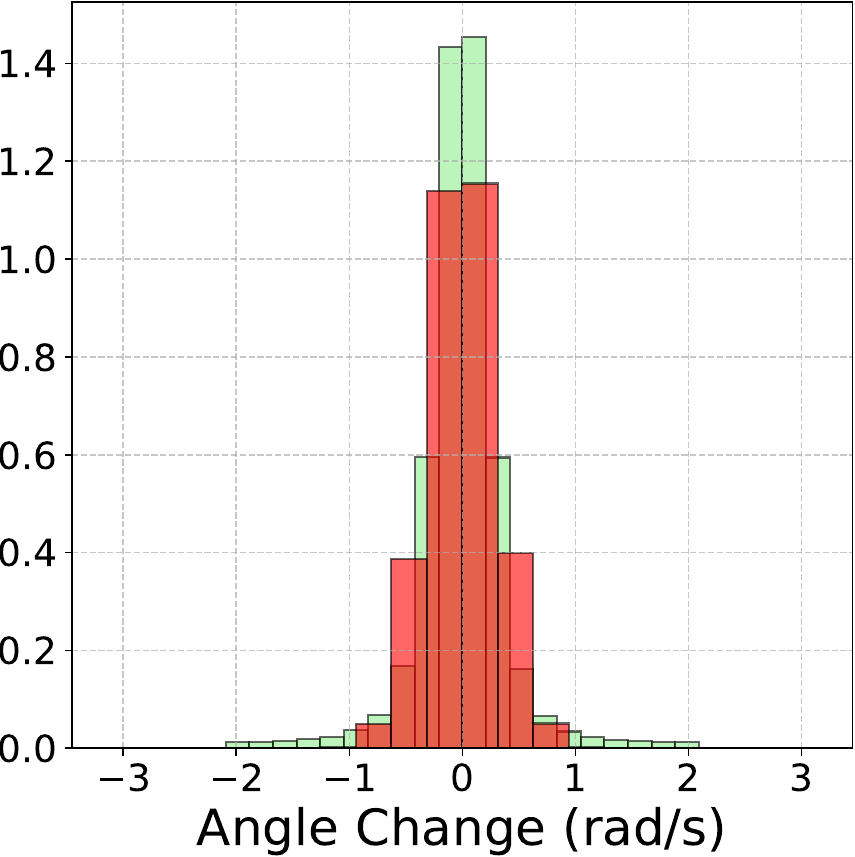} &
        \includegraphics[width=\linewidth/7]{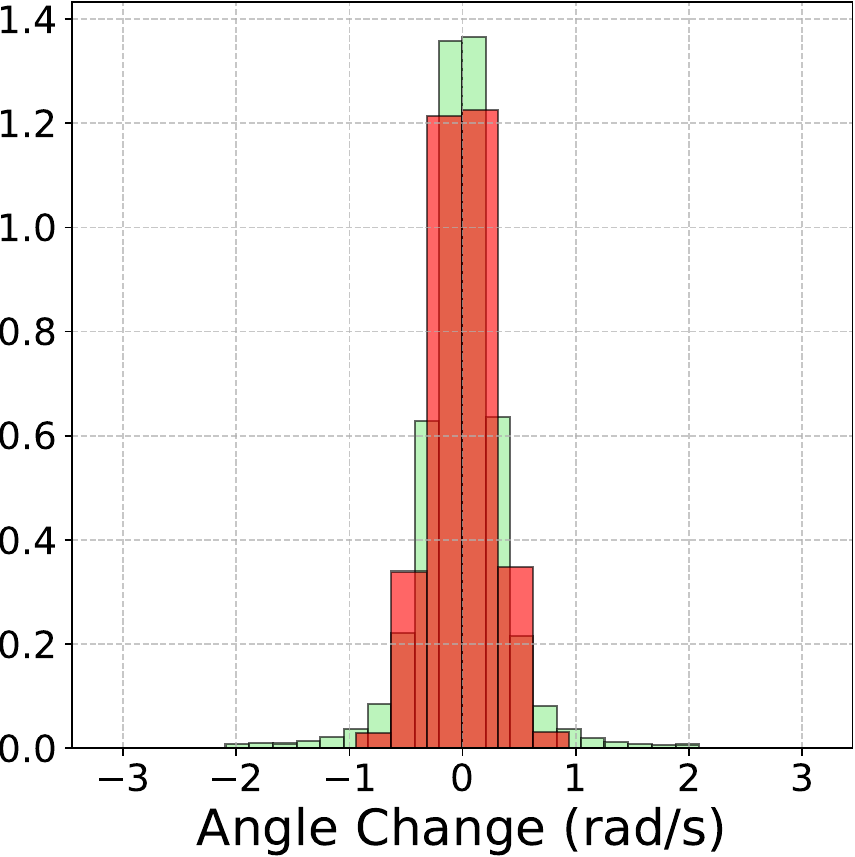} &
        \includegraphics[width=\linewidth/7]{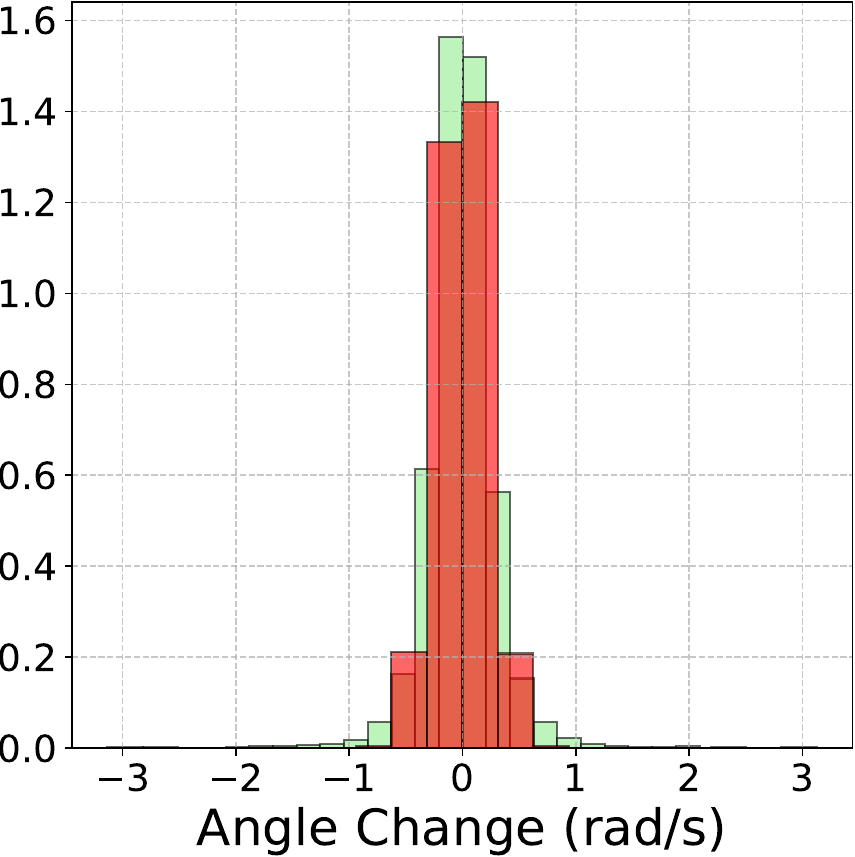} \\

        (a) & (b) & (c) & (d) & (e) & (f) \\   
    \end{tabular}
    
    \vspace{1em}

    \begin{tabular}{*{5}{c}} 
        \includegraphics[width=\linewidth/7]{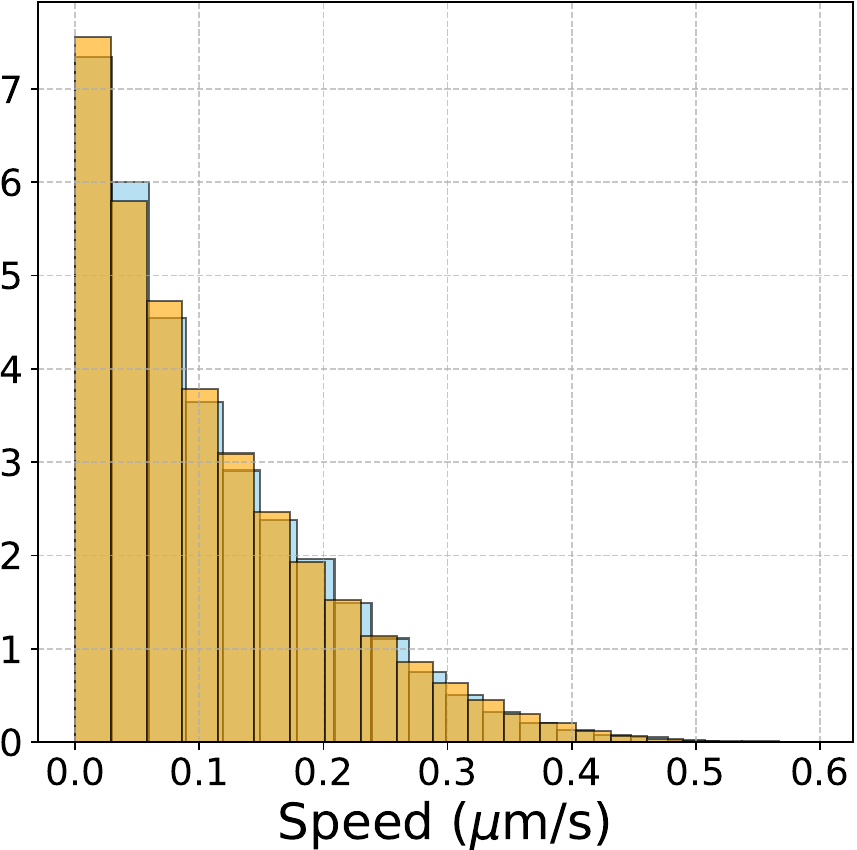} &
        \includegraphics[width=\linewidth/7]{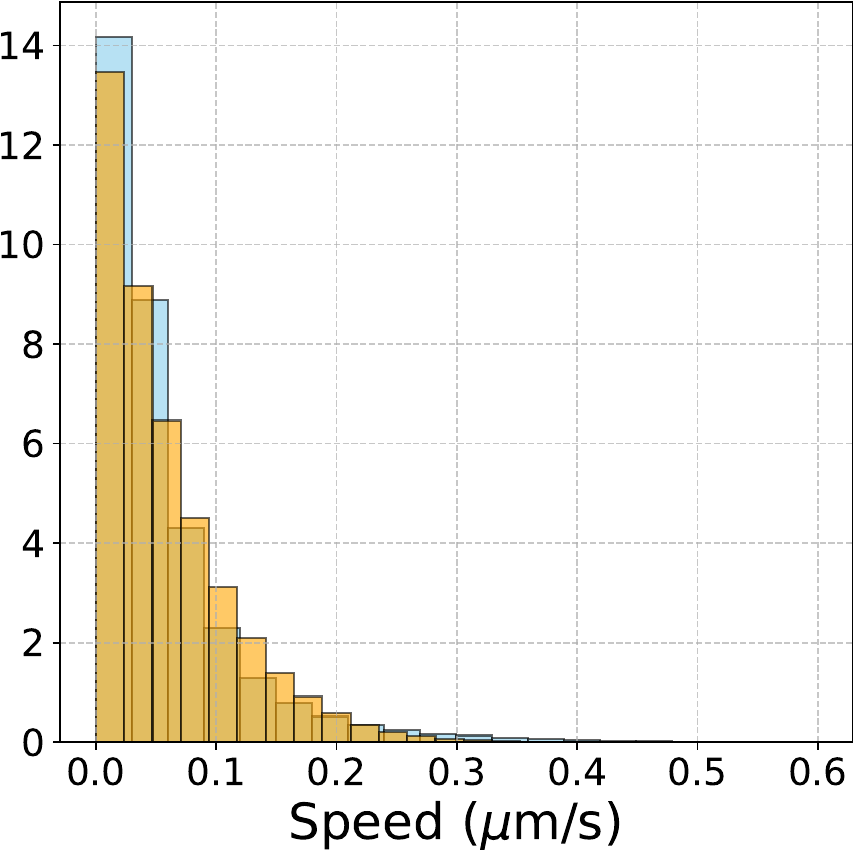} &
        \includegraphics[width=\linewidth/7]{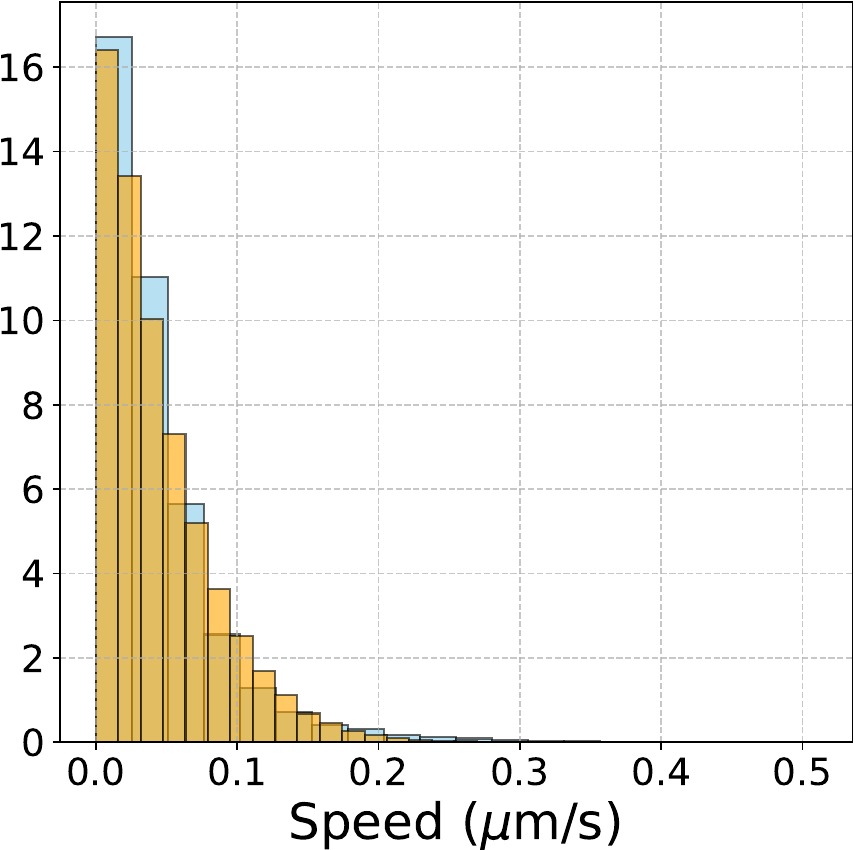} &
        \includegraphics[width=\linewidth/7]{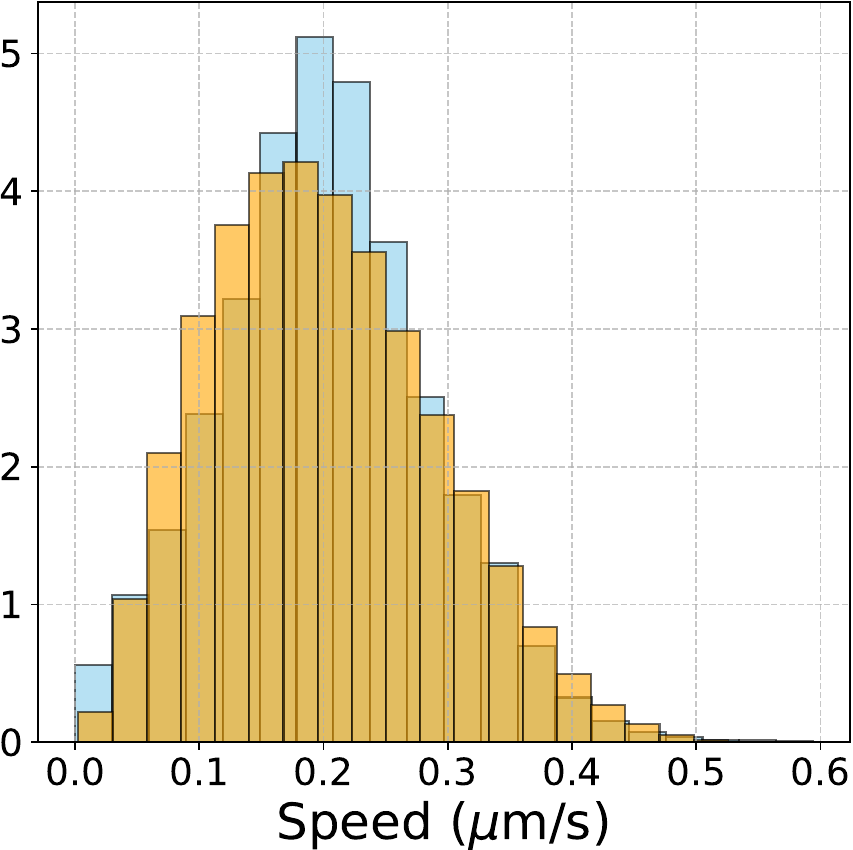} &
        \includegraphics[width=\linewidth/7]{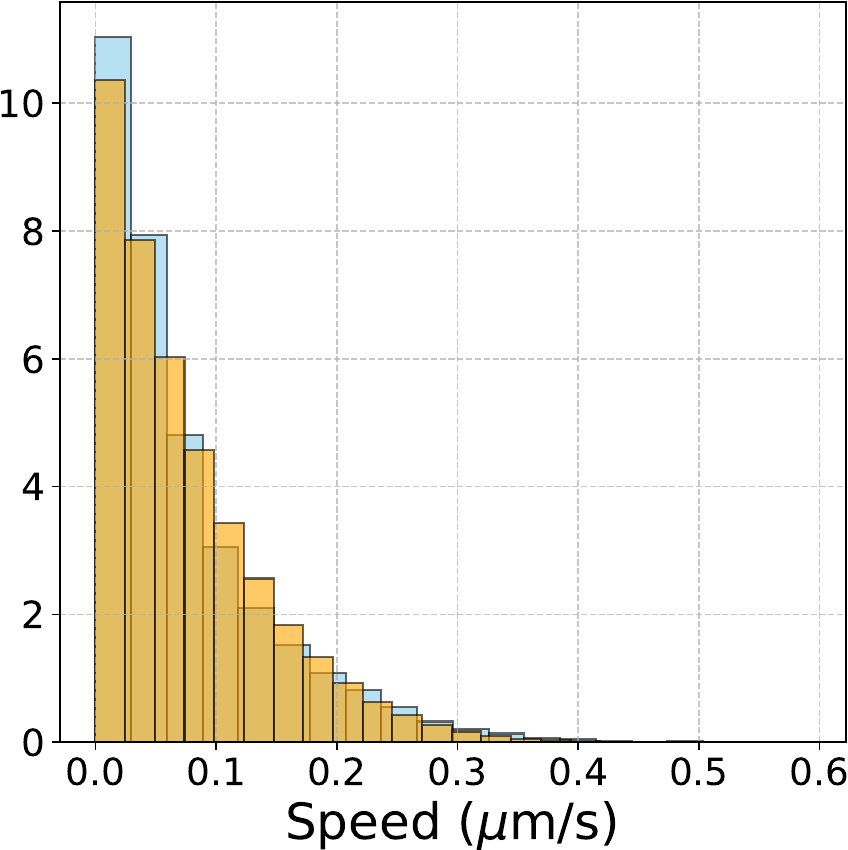} \\
        
        \includegraphics[width=\linewidth/7]{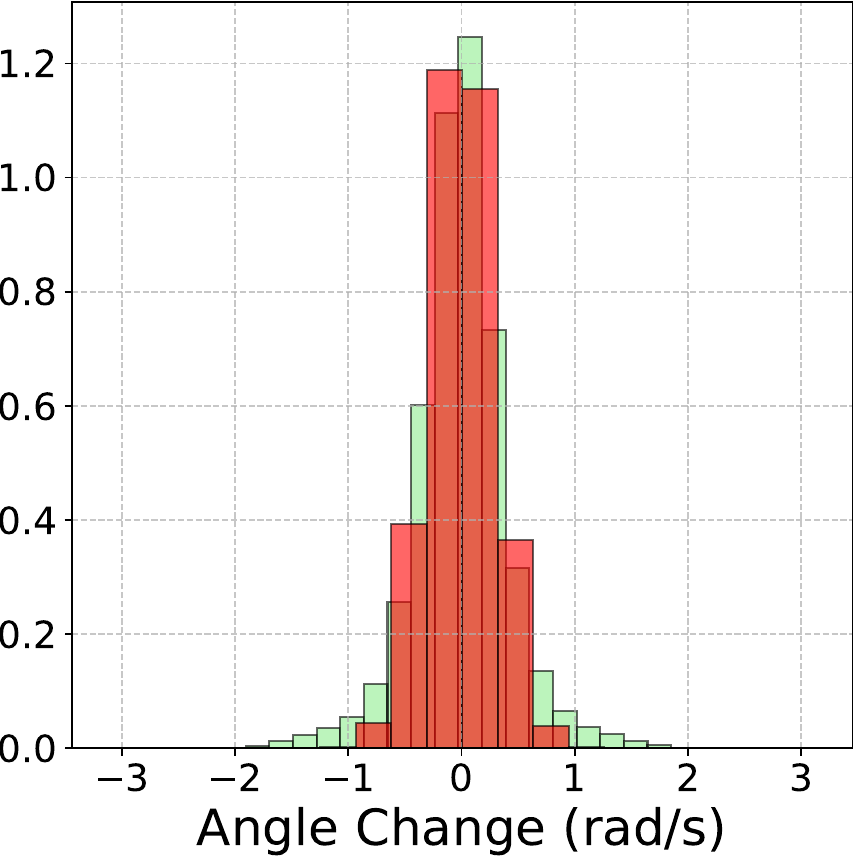} &
        \includegraphics[width=\linewidth/7]{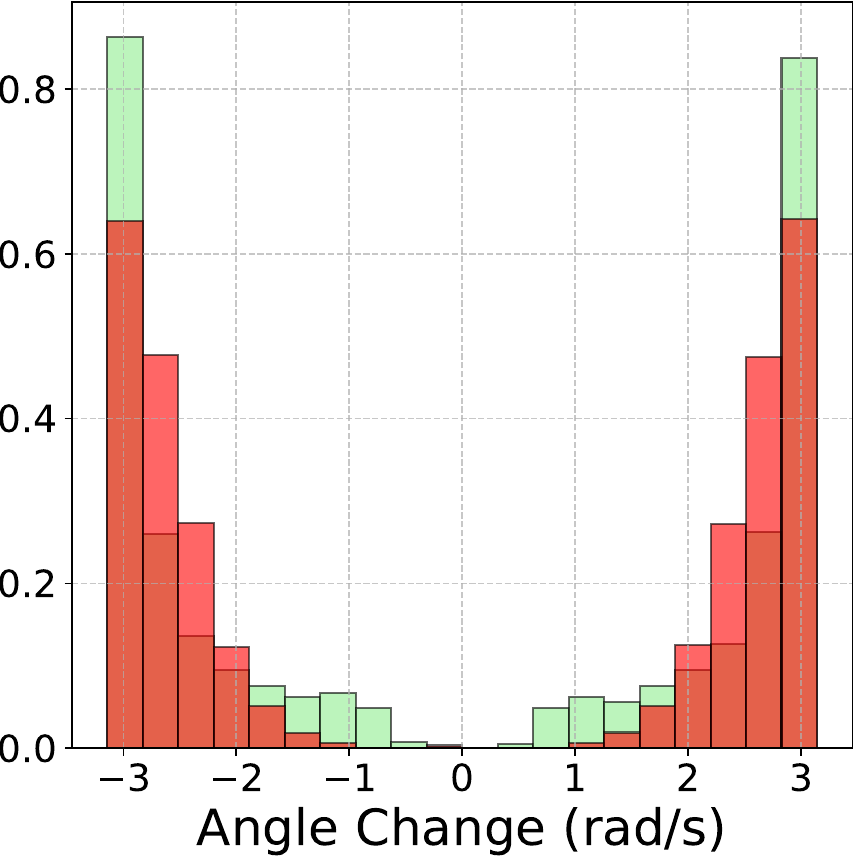} &
        \includegraphics[width=\linewidth/7]{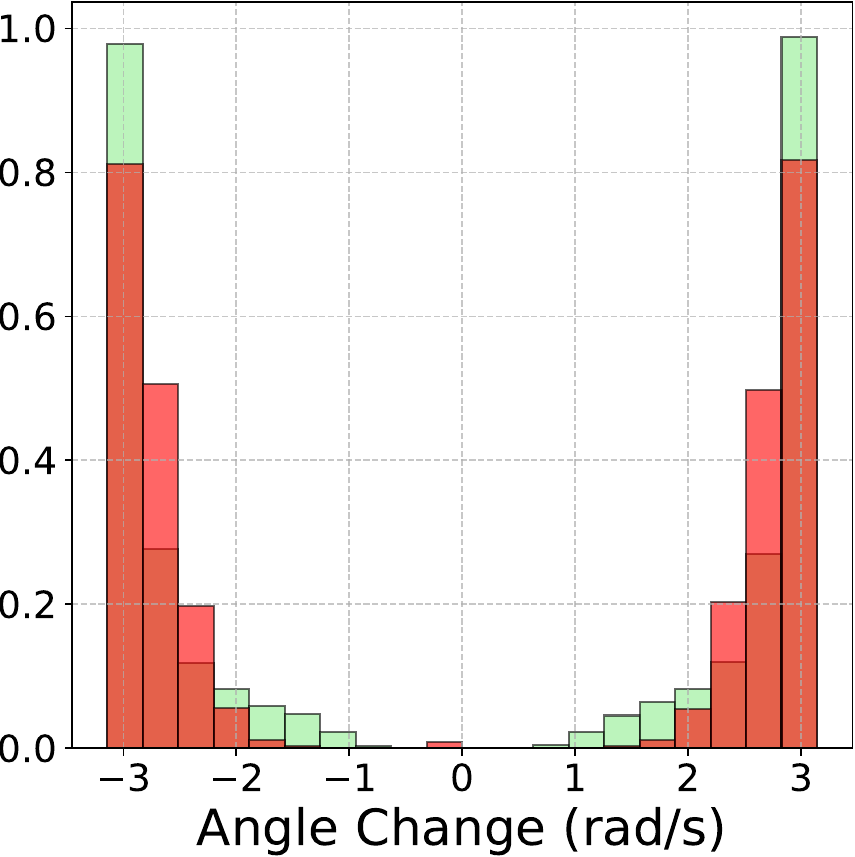} &
        \includegraphics[width=\linewidth/7]{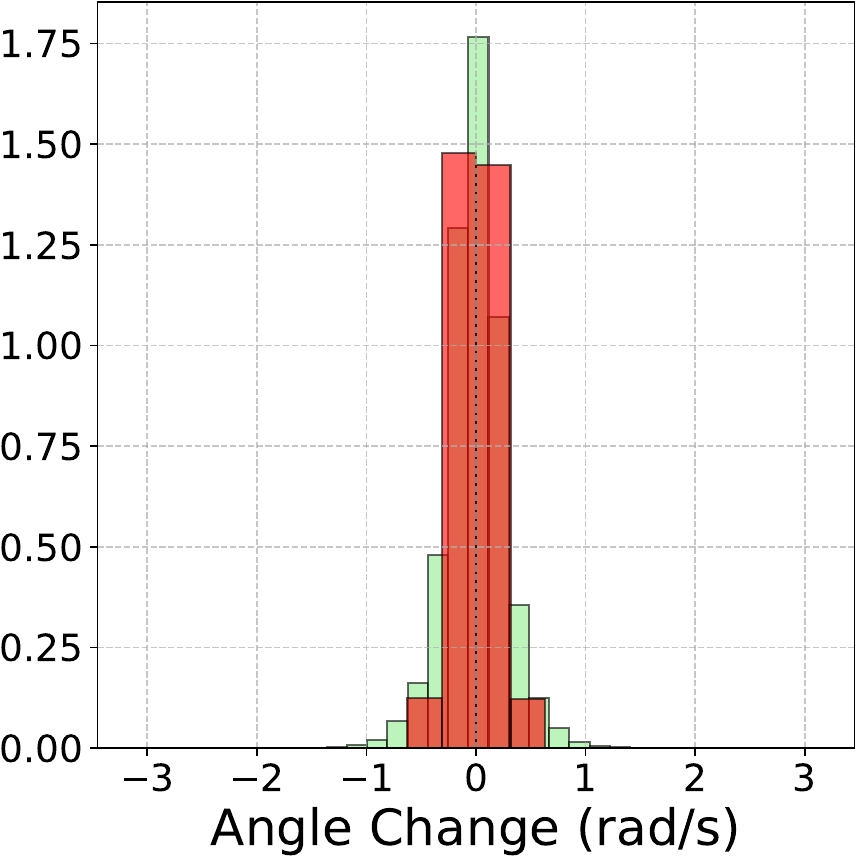} &
        \includegraphics[width=\linewidth/7]{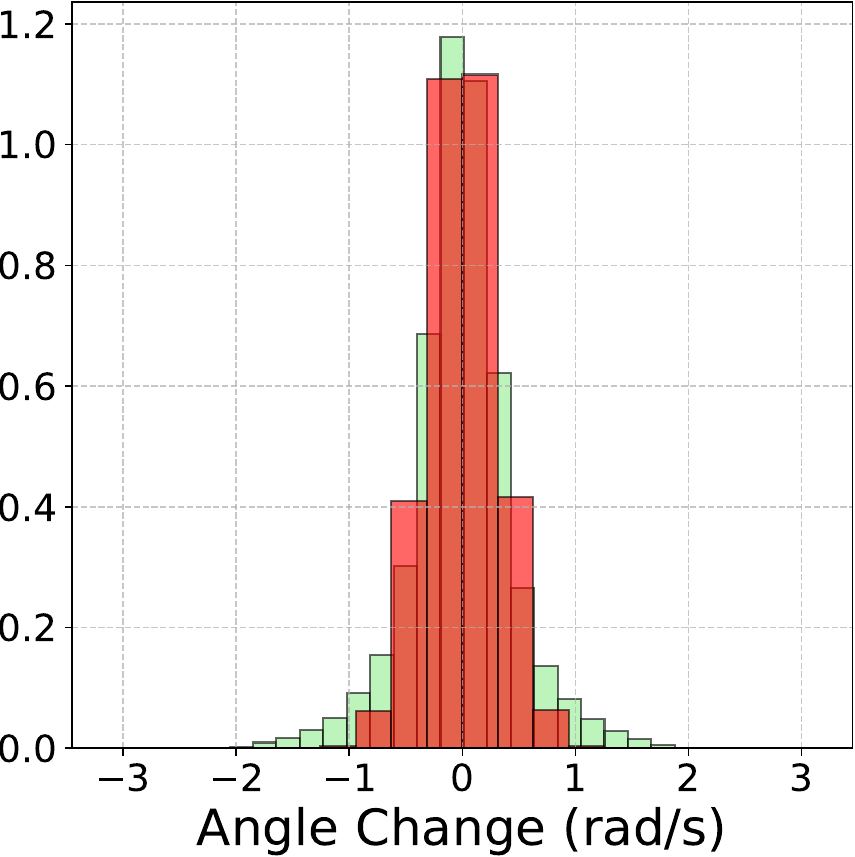} \\
        (g) & (h) & (i) & (j) & (k) \\
    \end{tabular}

    \caption{Distributions of speed (skyblue) and angle change (light green) for each atomic behaviour (top) and each automatic behaviour (bottom) when classifying our dataset. We overlay the distributions obtained from our agent-based simulator when using $1000$ agents for $1800$ time steps (speed in orange, angle change in red) and report the Kolmogorov-Smirnoff (KS) statistic in Table~\ref{tab:ks_atomic_and_automatic}. For the atomic behaviours, speed and angle change are computed for state (a) sharp turn, (b) reversal, (c) pause, (d) lines, (e) arcs and (f) loops. For the automatically defined behaviours, speed and angle changes are computed for states (g) slow-line ($0$), (h) straight-turn ($1$), (i) loop-turn ($2$), (j) crawl ($3$) and (k) high-turning ($4$).}
    \label{fig:models_histograms}
\end{figure*}

We begin by analysing the distributions of the worm speed and angular change in each state when classifying through the method described in Section~\ref{sec:atomic_behaviour}. The resulting distributions are shown in Figure~\ref{fig:models_histograms} (top), where sharp turns are characterised by a threshold on the angle change, thus only points with angle changes of an absolute value above $\frac{2}{3}\pi$ are included as per Figure~\ref{fig:models_histograms}a. Similarly, pauses are characterised by a threshold on the speed, with only points where the speed is below $50\mu m s^{-1}$ as in Figure~\ref{fig:models_histograms}c. The crawling states are characterised by: (i) an angle change distribution which has an varying $\kappa$ when modelled as a Von Mises distribution (higher for lines, lower for loops) (Figures~\ref{fig:models_histograms}d-f); (ii) a generally higher speed for the loop state than the other two crawl states (Figure~\ref{fig:models_histograms}f).

We then perform the same analysis on the distributions resulting from the automatic method described in Section~\ref{sec:automatic_behaviour}. The resulting distributions are shown in Figure~\ref{fig:models_histograms} (bottom). We notice that states $1$ and $2$ are characterised by similar distributions of angle changes and speed, as in Figures~\ref{fig:models_histograms}h-i. Similarly, the distribution of angle changes is comparable between states $0$ and $4$, while the speed distribution of the former are slightly more skewed towards lower values when compared to the latter (see Figure~\ref{fig:models_histograms}g and Figure~\ref{fig:models_histograms}k). On the other hand, state $3$ is skewed towards relatively higher speeds, suggesting that state $3$ is a crawling state, encompassing all forward movement (Figure~\ref{fig:models_histograms}j).  

\begin{table}[]
    \centering
    \begin{tabular}{lcc}
        \toprule
        \textbf{Atomic classification} & \textbf{Speed KS} & \textbf{Heading KS} \\
        \midrule
        Sharp turn & 0.078 & 0.072 \\
        Reversal   & 0.041 & 0.047 \\
        Pause      & 0.039 & 0.065 \\
        Line       & 0.030 & 0.042 \\
        Arc        & 0.060 & 0.031 \\
        Loop       & 0.062 & 0.031 \\
        \bottomrule
        \textbf{Automatic classification} & \textbf{Speed KS} & \textbf{Heading KS} \\
        \midrule
        Slow-line     & 0.078 & 0.070 \\
        Straight-turn & 0.069 & 0.080 \\
        Loop-turn     & 0.050 & 0.078 \\
        Crawl         & 0.058 & 0.060 \\
        High-turning  & 0.078 & 0.073 \\
        \bottomrule
    \end{tabular}
    \caption{KS distances between distributions of $1000$ simulated and real worm features for each atomic and automatic behavioural state.}
    \label{tab:ks_atomic_and_automatic}
\end{table}

\begin{table}[]
    \centering
    \begin{tabular}{lrrrr}
    \toprule
    \textbf{Class} & \textbf{Precision} & \textbf{Recall} & \textbf{F1-score} & \textbf{Support} \\
    \midrule
    0 (slow-line) & 0.98 & 0.98 & 0.98 & 6162 \\
    1 (straight-turn) & 0.98 & 0.98 & 0.98 & 4078 \\
    2 (loop-turn)& 0.98 & 0.98 & 0.98 & 2812 \\
    3 (crawl) & 0.99 & 0.99 & 0.99 & 6766 \\
    4 (high-turning) & 0.98 & 0.98 & 0.98 & 4399 \\

    \midrule
    \textbf{Accuracy} & & & \textbf{0.98} & \textbf{24217} \\
    \textbf{Macro avg} & 0.98 & 0.98 & 0.98 & 24217 \\
    \textbf{Weighted avg} & 0.98 & 0.98 & 0.98 & 24217 \\
    \bottomrule
    \end{tabular}
    \caption{Classification report with precision, recall, and F1-score for each class for the XGBoost classifier for the prediction of automatically defined behavioural states from kinematic, spatial and temporal features. Overall accuracy and macro/weighted averages are also reported.}
    \label{tab:classification_report}
\end{table}
\begin{figure*}[t]
    \centering

        \begin{tabular}{ccc}
 
        \includegraphics[width=0.32\linewidth]{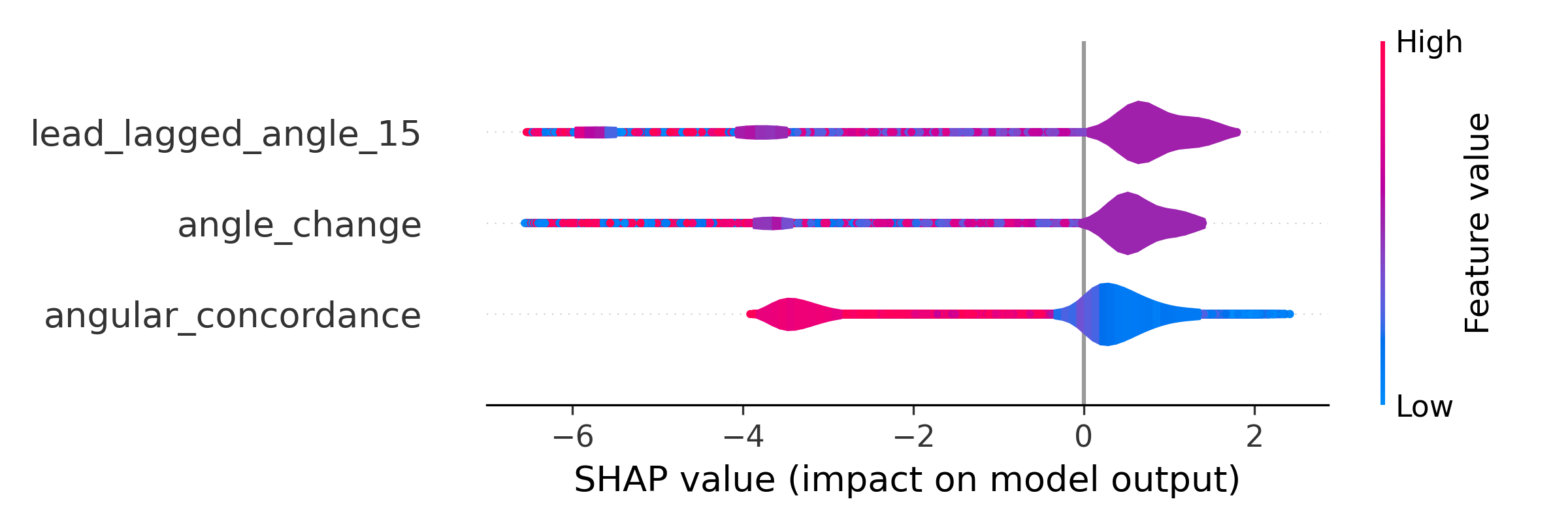}  & \includegraphics[width=0.32\linewidth]{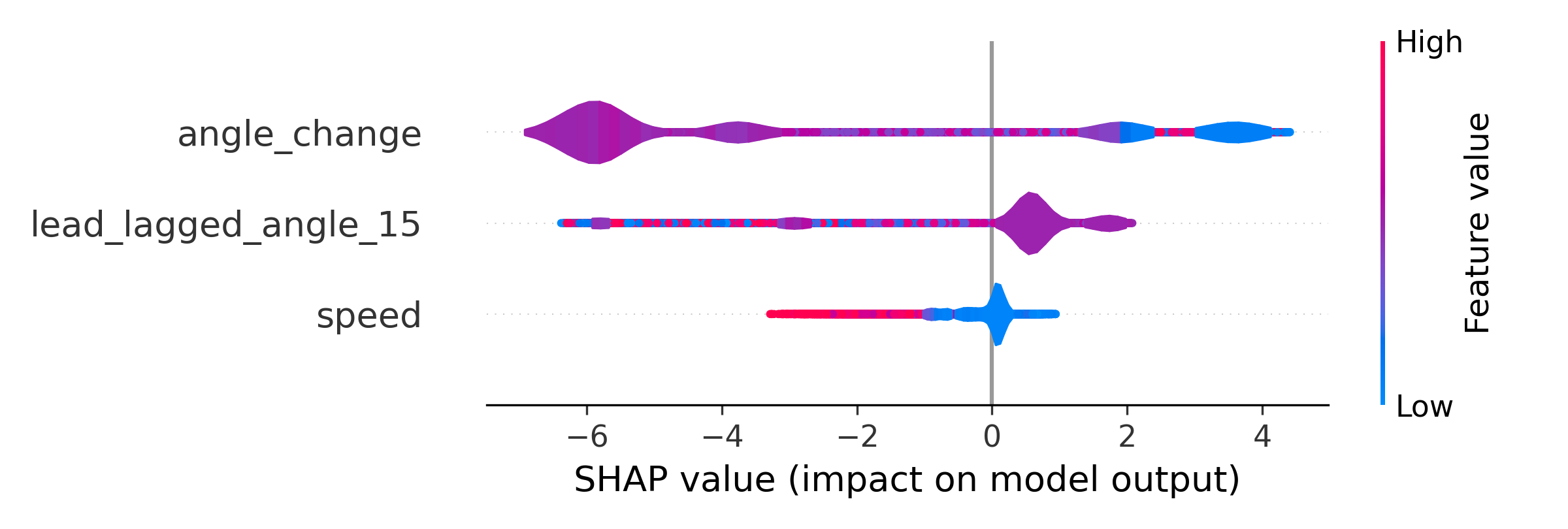}  & 
        \includegraphics[width=0.32\linewidth]{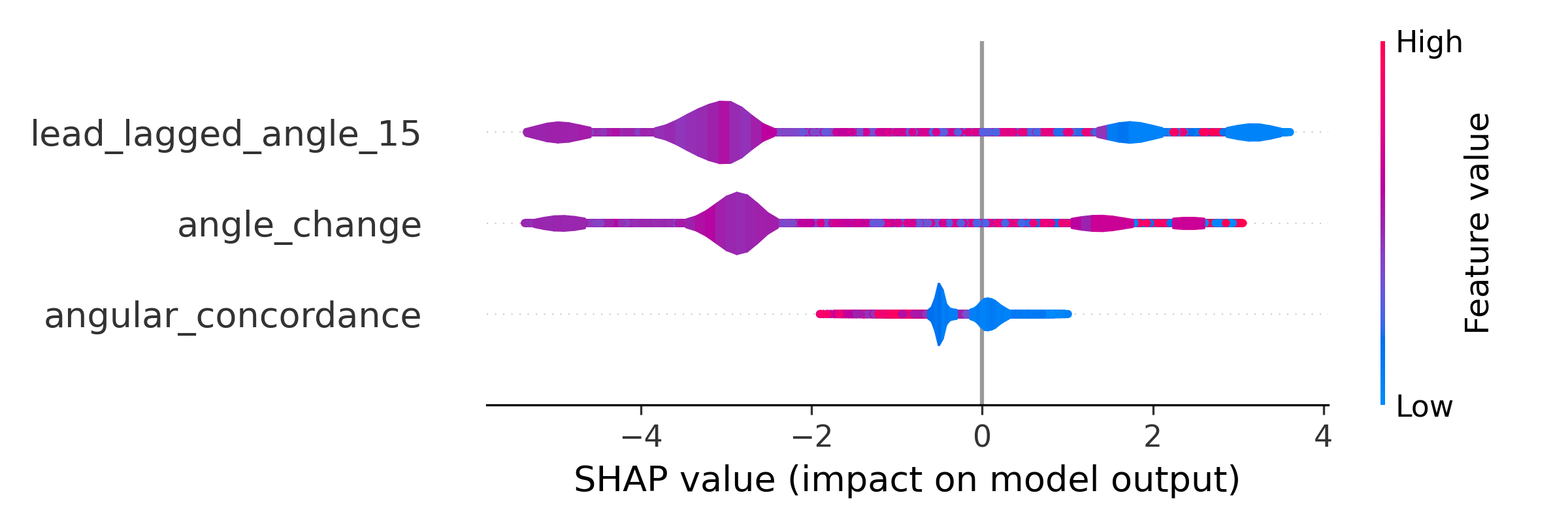}  \\
        (a) & (b) & (c) \\
        \includegraphics[width=0.32\linewidth]{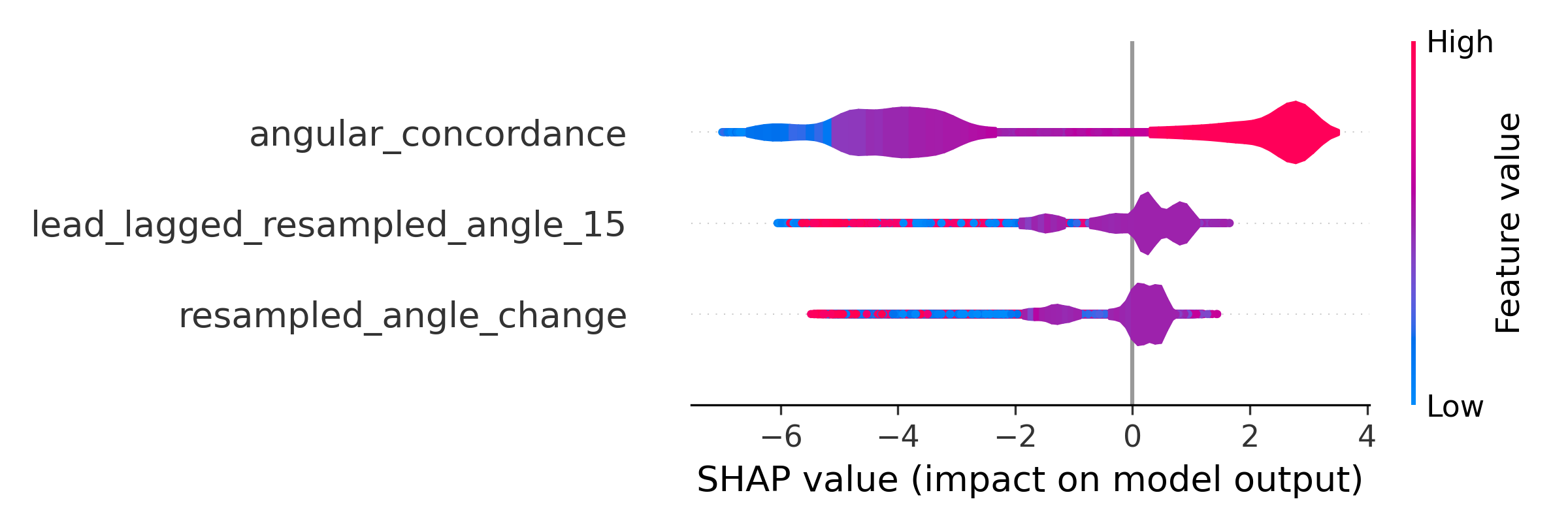}   &
         \includegraphics[width=0.32\linewidth]{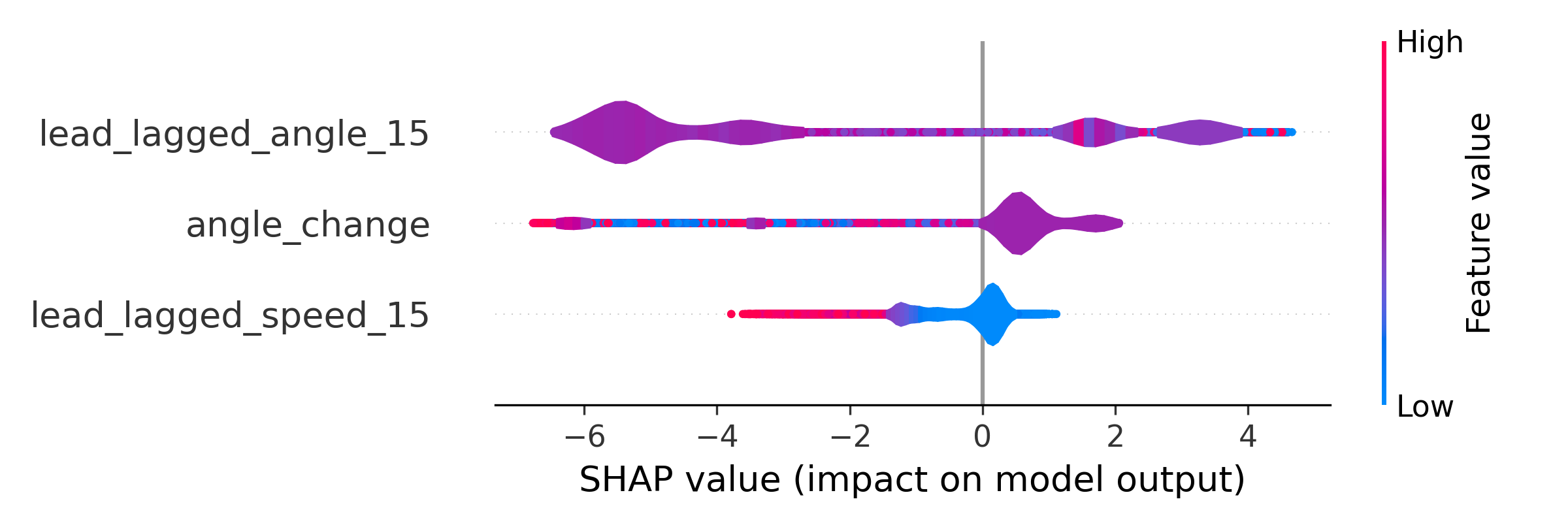} \\
        (d) & (e) \\              
        \end{tabular}

    \caption{Shapley values of the features for each state obtained from the automatic behaviour method after an optimised XGBoost predictor was trained on the features to classify the state labels after the KMeans clustering on the embedded feature space with $k^*=5$ clusters. We show only the $3$ most influent Shapley values for state (a) slow-line ($0$), (b) straight-turn ($1$), (c) loop-turn ($2$), (d) crawl ($3$) and (e) high-turning ($4$). }
    \label{fig:shap}
\end{figure*}

In order to gain further insights into the properties of the automatically obtained states, we train an XGBoost predictor with a stratified group 5-Fold cross validation on the features against the predicted states for each data point and tune its hyperparameters via Bayesian optimisation. The resulting classification report is in Table~\ref{tab:classification_report}. Then, we find the Shapley values of the features for each state as predicted by XGBoost and plot them in Figure~\ref{fig:shap}. By combining the latter with Figure~\ref{fig:models_histograms}, we find that state $0$ and state $4$ are characterised by short and mostly straight runs. The two differ in the lagged angle changes: for state $0$, it is centred around $0$rad/s (Figure~\ref{fig:shap}a), while for state $4$, it is centred at extreme values ($\pm \pi$ rad/s, Figure~\ref{fig:shap}e). Given this difference, we name state $0$ \textit{slow-line} and state $4$ \textit{high-turning}, given the time-correlated high turning angle. On the other hand, states $1$ and $2$ both are composed of sharp turns and lower speed, however the former is correlated to straight runs, as can be seen from its lagged angle change (Figure~\ref{fig:shap}b), while the latter has lagged angle changes that tend to extreme values ($\pm \pi$ rad/s, Figure~\ref{fig:shap}c). Therefore, in the following, we define state $1$ as \textit{straight-turn} and state $2$ as \textit{loop-turn}. State $3$ is characterised by longer durations and relatively high speed, with angle changes mostly concentrated around $0$rad/s (Figure~\ref{fig:shap}d), all properties compatible with crawls characterised by an arc or loop trajectory, thus we define it as \textit{crawl}. These findings show that the clustering pipeline tends to prioritise temporal dependencies, especially of the angle change, rather than instantaneous changes in direction of movement, suggesting that temporally correlated patterns exist when treating the worm as a point. 

\begin{figure}[t]
    \centering
    \includegraphics[width=0.66\linewidth]{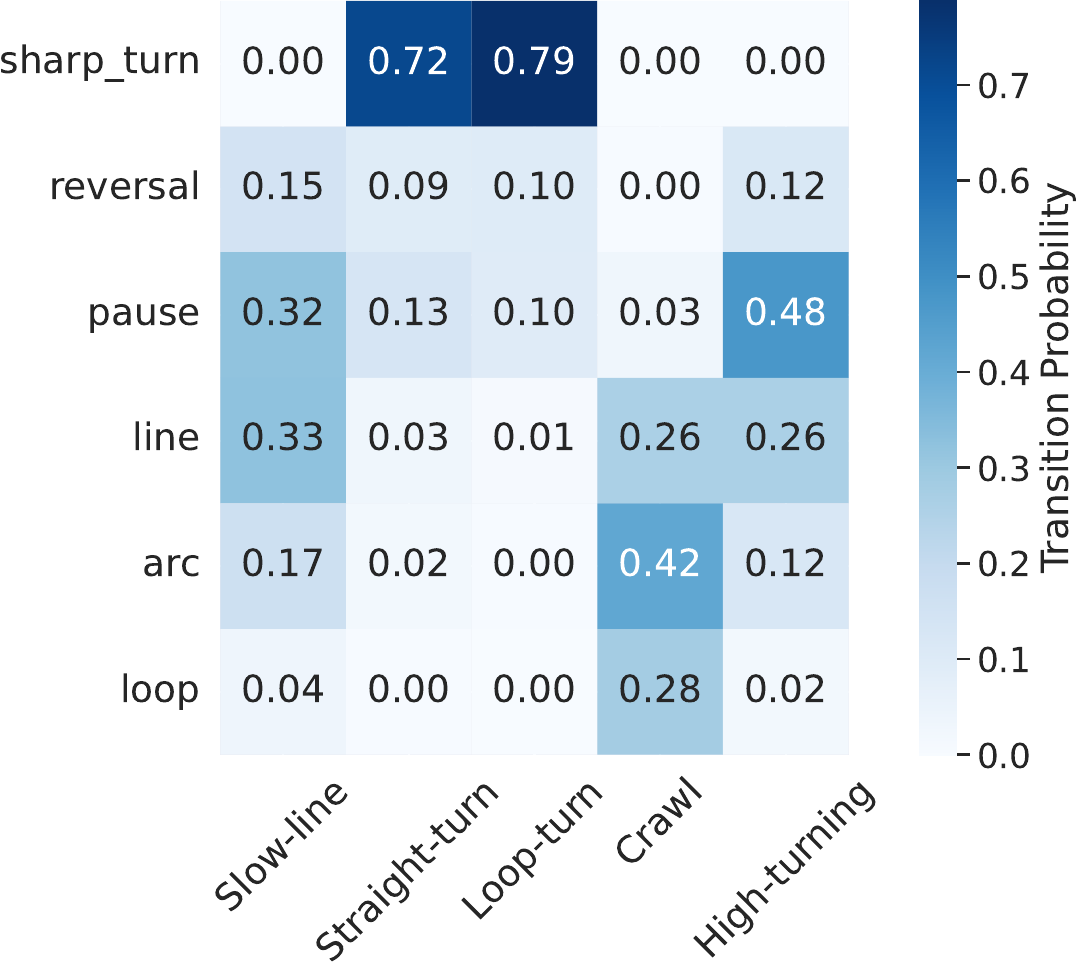}
    \caption{Ordered confusion matrix of the two classification methods: rows represent labels obtained by applying the atomic method, whereas columns represent labels obtained by the automatic classification. Each cell contains the number of data points through time where the two systems agree and is normalised column-wise, thus showing how each automatic state maps to the atomic states.}
    \label{fig:confusion_matrix}
\end{figure}

We analyse the agreement of the two labelling systems by defining the ordered confusion matrix which has a number of rows equal to the number of atomic behaviours ($6$) and a number of columns equal to the number of automatically detected behaviours ($5$). Each cell $i, j$ contains the count of data points which, at the same time step, are labelled as $i$ and as $j$ by the atomic and automatic systems, respectively, and normalised with respect to the number of occurrences of label $j$. The obtained confusion matrix is shown in Figure~\ref{fig:confusion_matrix} and indicates that our atomic division of behavioural states emerges partially from the automatic pipeline. In particular, reversals are not categorised into a state, even though the resampled angle change with respect to body length was included in the features. This result is expected as the emergence of reversals is not trivial when the worm is reduced to a point. On the other hand, states straight-turn and loop-turn are almost entirely comprised of sharp turns, in agreement with our previous analysis where we show that they differ in the lagged angle change (as shown in Figure~\ref{fig:shap}b and Figure~\ref{fig:shap}c). Moreover, states slow-line and high-turning emerge as a combination of pauses and line crawls, while differing in the lagged angle change (see Figure~\ref{fig:shap}a and Figure~\ref{fig:shap}e). Finally, state crawl is composed of straight (lines) and curved trajectories (arcs and loops), representing a generic crawling state and suggesting that arcs, lines and loops might be condensed into a single state where the key difference is the distribution of angle changes. 
 
\subsection{Performance} 
\label{sec:performance}
We assess the performance of the atomic and the automatic methods by constructing the agent-based model described in Section~\ref{sec:ab-model}.
First, we analyse the number of behavioural events per $120$s under the two methods in order to parametrise $m_i, q_i$ in Equation~\ref{eq:transition_prob}. In contrast to~\citep{Salvador2014}, we find no time-dependent trends in the count of behavioural events in neither the atomic nor in the automatic classification methods, as shown by the overall near $0$ slope ($m$) of the line fitting on the mean number of behavioural events in Table~\ref{tab:count_line_fit_parameters} for both classification methods. This result is expected as our dataset is composed of tracks of worms on food, while~\citep{Salvador2014} use a dataset of worms without food, a condition that leads to a switch in the worm behaviour from a local search to a global search strategy. Such modulation is absent in worms feeding~\citep{bonnard2022}, thus the mean number of events per time window remains mostly constant across time. 

\begin{table}[]
    \centering
    \begin{tabular}{l|c|c||l|c|c}
    \multicolumn{3}{c||}{\textbf{Atomic classification}} & \multicolumn{3}{c}{\textbf{Automatic classification}} \\
    \hline
    State      & m     & q     & State           & m     & q     \\
    \hline
    Sharp turn & -0.01 & 1.20  & Slow-line       & 0.02  & 1.32  \\
    Reversal   & 0     & 0.50  & Straight-turn   & 0.01  & 1.14  \\
    Pause      & -0.01 & 1.41  & Loop-turn       & -0.02 & 0.65  \\
    Line       & 0.02  & 0.72  & Crawl           & 0.02  & 0.31  \\
    Arc        & 0.01  & 0.28  & High-turning       & 0.01  & 1.19  \\
    Loop       & 0     & 0.02  &                 &       &       \\
    \end{tabular}
    \caption{Parameters obtained by fitting a line on the mean number of behavioural events per $120$s window for each state under the atomic and automatic classification methods. }
    \label{tab:count_line_fit_parameters}
\end{table}

\begin{table}[]
    \centering
    \begin{tabular}{@{}lcr@{}}
        \toprule
        \textbf{Method} & \textbf{Agents} & \textbf{Log-likelihood} \\
        \midrule
        Atomic     & 100           &  -41.14 \\
        Atomic     & 1000           & -26.31 \\
        Automatic  & 100           & \textbf{-21.64} \\
        Automatic  & 1000   & \textbf{-22.09} \\
        \bottomrule
    \end{tabular}
    \caption{
        Model performance across methods and agent counts, measured by the mean log-likelihood of the mean square displacement. Best performance (maximum log-likelihood) is achieved using automatically defined behaviours.
    }
    \label{tab:performance}
\end{table}

\begin{figure}[t]
    \centering
    \begin{tabular}{cc}
        \includegraphics[width=0.49\linewidth]{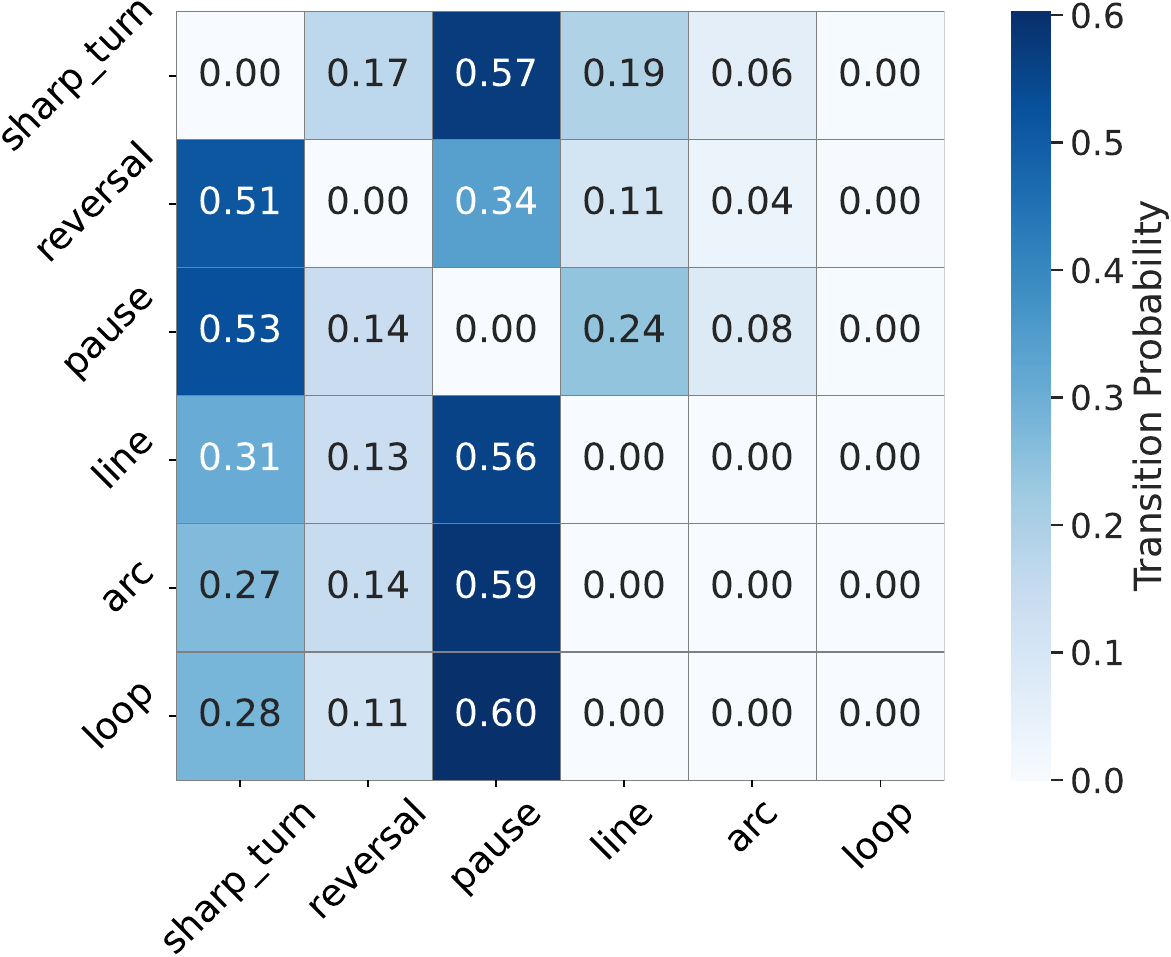} &
        \includegraphics[width=0.49\linewidth]{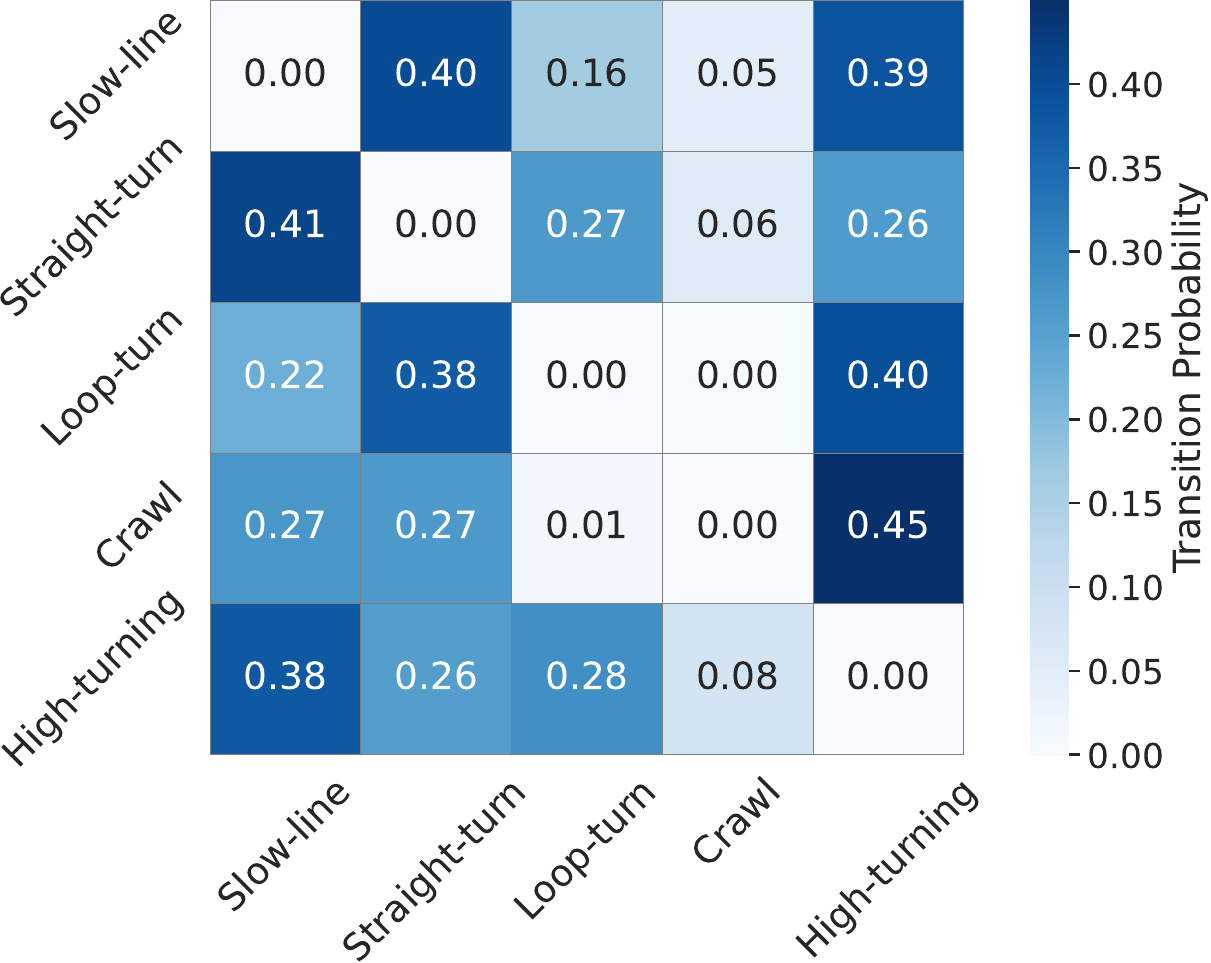} \\
        (a) &(b) \\
    \end{tabular}

    \caption{
        Transition matrices for the atomic (a) and automatic (b) methods computed as the overall count of transitions between behavioural states, excluding self-transitions and normalised row-wise.
    }
    \label{fig:counts_and_trans}
\end{figure}

We compute the transition matrix $l_{s\rightarrow s'}$ from Equation~\ref{eq:transition_prob} by counting the transitions between states in the two classification methods. In Figure~\ref{fig:counts_and_trans}a we show the transition matrix for the atomic behaviours, where crawling states and sharp turns have a higher probability of leading to a pause, while reversals and pauses have a positive correlation with sharp turns.
The transition matrix for the automatically defined behaviours in Figure~\ref{fig:counts_and_trans}b suggests that slow-line, loop-turn and crawl states lead to the high-turning state most often, while the straight-turn and high-turning states lead most often to the slow-line state.

\begin{figure*}[ht]
    \centering
    \begin{tabular}{cc}
    \includegraphics[width=0.49\textwidth]{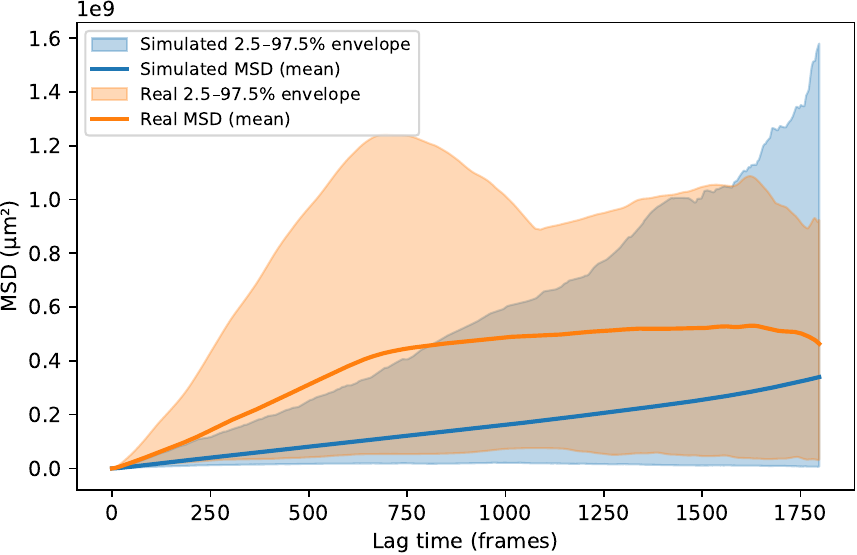}     &  \includegraphics[width=0.49\textwidth]{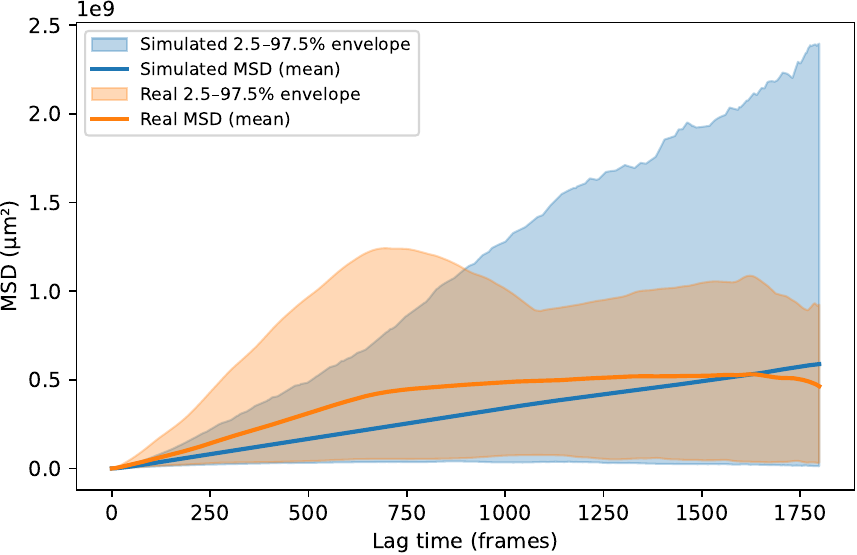}\\
     (a)    & (b) \\
    \end{tabular}
    \caption{ Mean square displacement of biological worms and 1000 simulated agents using the atomic (a) and automatic (b) behavioural classification computed on lags of up to $1800$s.}
    \label{fig:msd}
\end{figure*}

Next, we compare the performance of the two classification methods by simulating two probabilistic finite-state machines (Section~\ref{sec:ab-model}), one for each method, with $100$ and $1000$ agents. The resulting distributions of speed and angle change for each state of $1000$ simulated agents are shown in Figure~\ref{fig:models_histograms} (top atomic, bottom automatic). As can be seen from the Kolmogorov-Smirnof distance metric in Table~\ref{tab:ks_atomic_and_automatic}, the maximum difference between the simulated and real histograms never exceeds $7.8\%$, indicating that our fitted distributions are in line with the distributions of biological worms across all behavioural states. Furthermore, we compute the distribution of mean square displacement (MSD) at each time lag $\tau \in [1, 1800]$ for the simulated agents and the biological worms. We measure the average log-likelihood at lag $\tau$ with method $i$ and number of agents $j$ as:
\begin{equation}
    l_{i,j}(\tau) = \frac{1}{N}\sum_{w=1}^N\log \left[  \mathcal{N}\left(\text{MSD}_w(\tau) | \mu_{i,j}(\tau), \sigma_{i,j}(\tau)\right) \right]
\end{equation}

where $\text{MSD}_w(\tau)$ is the MSD of worm $w$ for lag $\tau$ across all worms, $ \mathcal{N}(x | \mu_{i,j}(\tau), \sigma_{i,j}(\tau))$ is the probability density function of a normal distribution with mean $\mu_{i,j}(\tau)$ and standard deviation $ \sigma_{i,j}(\tau)$ and $N$ is the number of worm tracks with a duration of at least $1800$s. We report the average log-likelihood in Table~\ref{tab:performance}, where $l_{i,j}$ increases as $j$ increases and the maximum values are achieved when using the automatic classification method. The result is in line with the distribution of MSD across $\tau$ in Figure~\ref{fig:msd} computed for $1000$ agents for the atomic (Figure~\ref{fig:msd}a) and the automatic (Figure~\ref{fig:msd}b) methods, suggesting that by leveraging temporal dependencies, the automatic model is able to naturally capture behavioural correlations, which are overlooked by the atomic model. Both methods tend to undershoot the mean MSD of real worms, however the automatic method is characterised by a higher spread of MSD as $\tau$ increases.


\section{Conclusion}
\label{sec:conclusion}
In this paper, we compare two methods to characterise the behaviour of a \celeg\ worm when tracking only its body centre point. 
Existing methods for the analysis of the worm behaviour require the view of the full worm body during tracking in order to segment its body centre line.  
Contrary to this literature, we introduce a novel approach based on tracking and analysing a single worm body point moving in a $2$D space, thus facilitating the tracking of worms also in high worm density conditions, in order to capture the effects of the social environment on single worm movements. 

Our main contribution consists in the evaluation of the performance of hand-designed atomic behaviours against a fully automated behaviour classification pipeline. 
We build an agent-based model which, albeit simple, is capable of reproducing similar trends in movement to that of biological worms. Our results show that the automatic method, which takes into account temporal dependencies between movement trends, performs better than the atomic method when using our agent-based model, suggesting that sufficient temporal correlations exist to determine behavioural states even at the scale of a point. 

Despite our best efforts, our analysis is still preliminary. First of all, the definition of the atomic behaviours is arbitrary and the exclusion of temporal dependencies might undermine the atomic behaviour performance. On the other hand, the inclusion of temporal dependencies clashes with the intention of defining atomic units of behaviour. Moreover, the agent-based model is built with temporal dependencies on the transitions between states exactly to overcome this limitation. Albeit simple, our agent-based model captures first order temporal dependencies, which previous work~\citep{Costa2024MarkovWorm} showed to be sufficient for modelling \celeg\ behaviour. Our results indicate that more attention should be devoted to the analysis of automated behaviour recognition pipelines where lagged features are absent. These studies should compare the performance of the automated behaviour lacking lagged features against that of the atomic methods, in order to assess whether lagged features might add a second order temporal dependency.

Another property that bears upon the effectiveness of the automatic behaviour classification method is the dependence of the clustering on the embeddings on the original data. By removing or adding data points the optimal clustering and its properties might vary, thus altering the definition of the behavioural states. This becomes of paramount importance when comparing the two methods: our data is from worms feeding, where the majority of the time the worms are pausing to feed. Perhaps, in another condition, the data might get skewed and other trends might arise, obscuring already existing clusters or generating novel ones. For example, in the presence of aversive stimuli, the reversal might emerge as a behaviour simply given its higher frequency. Thus, future work should also focus on the comparison of multiple worms conditions in order to assess how stable the optimal clustering is, together with its behavioural building blocks.

\section{Acknowledgements}
The BABots project has received funding from the Horizon Europe, PathFinder European Innovation Council Work Programme under grant agreement No 101098722. Views and opinions expressed are however those of the authors only and do not necessarily reflect those of the European Union or European Innovation Council and SMEs Executive Agency (EISMEA). Neither the European Union nor the granting authority can be held responsible for them.

\footnotesize
\bibliographystyle{apalike}
\bibliography{bibliography}

\begin{thebibliography}{}

\bibitem[BABots, 2024]{BABOTS_WebPage}
BABots (2024).
\newblock {BABots}: The design and control of small swarming biological animal robots.
\newblock \url{https://babots.eu/}.

\bibitem[Berman et~al., 2014]{berman2014}
Berman, G.~J., Choi, D.~M., Bialek, W., and Shaevitz, J.~W. (2014).
\newblock Mapping the stereotyped behaviour of freely moving fruit flies.
\newblock {\em Journal of The Royal Society Interface}, 11(99):20140672.

\bibitem[Bonnard et~al., 2022]{bonnard2022}
Bonnard, E., Liu, J., Zjacic, N., Alvarez, L., and Scholz, M. (2022).
\newblock Automatically tracking feeding behavior in populations of foraging c. elegans.
\newblock {\em Elife}, 11.

\bibitem[Broekmans et~al., 2016]{Broekmans2016}
Broekmans, O.~D., Rodgers, J.~B., Ryu, W.~S., and Stephens, G.~J. (2016).
\newblock Resolving coiled shapes reveals new reorientation behaviors in c. elegans.
\newblock {\em Elife}, 5.

\bibitem[Costa et~al., 2024]{Costa2024MarkovWorm}
Costa, A.~C., Ahamed, T., Jordan, D., and Stephens, G.~J. (2024).
\newblock A markovian dynamics for {{\it Caenorhabditis elegans}} behavior across scales.
\newblock {\em Proceedings of the National Academy of Sciences}, 121(32):e2318805121.

\bibitem[Eren et~al., 2024]{Eren2024}
Eren, G.~G., B{\"o}ger, L., Roca, M., Hiramatsu, F., Liu, J., Alvarez, L., Goetting, D., Zorn, N., Han, Z., Okumura, M., Scholz, M., and Lightfoot, J.~W. (2024).
\newblock Predatory aggression evolved through adaptations to noradrenergic circuits.
\newblock {\em bioRxiv}.

\bibitem[Hardaker et~al., 2001]{hardaker2001}
Hardaker, L.~A., Singer, E., Kerr, R., Zhou, G., and Schafer, W.~R. (2001).
\newblock Serotonin modulates locomotory behavior and coordinates egg-laying and movement in caenorhabditis elegans.
\newblock {\em Journal of Neurobiology}, 49(4):303--313.

\bibitem[Hsu and Yttri, 2021]{Hsu2021}
Hsu, A.~I. and Yttri, E.~A. (2021).
\newblock B-soid, an open-source unsupervised algorithm for identification and fast prediction of behaviors.
\newblock {\em Nature Communications}, 12(1):5188.

\bibitem[Javer et~al., 2018]{Javer2018}
Javer, A., Currie, M., Lee, C.~W., Hokanson, J., Li, K., Martineau, C.~N., Yemini, E., Grundy, L.~J., Li, C., Ch'ng, Q., Schafer, W.~R., Nollen, E. A.~A., Kerr, R., and Brown, A. E.~X. (2018).
\newblock An open-source platform for analyzing and sharing worm-behavior data.
\newblock {\em Nat Methods}, 15(9):645--646.

\bibitem[Liu et~al., 2018]{Liu2018}
Liu, M., Sharma, A.~K., Shaevitz, J.~W., and Leifer, A.~M. (2018).
\newblock Temporal processing and context dependency in \textit{Caenorhabditis elegans} response to mechanosensation.
\newblock {\em eLife}, 7:e36419.

\bibitem[McInnes et~al., 2018]{McInnes2018}
McInnes, L., Healy, J., Saul, N., and Großberger, L. (2018).
\newblock Umap: Uniform manifold approximation and projection.
\newblock {\em Journal of Open Source Software}, 3(29):861.

\bibitem[Salvador et~al., 2014]{Salvador2014}
Salvador, L. C.~M., Bartumeus, F., Levin, S.~A., and Ryu, W.~S. (2014).
\newblock Mechanistic analysis of the search behaviour of {{\it Caenorhabditis elegans}}.
\newblock {\em Journal of The Royal Society Interface}, 11(92):20131092.

\bibitem[Sarma et~al., 2018]{openworm2018}
Sarma, G.~P., Lee, C.~W., Portegys, T., Ghayoomie, V., Jacobs, T., Alicea, B., Cantarelli, M., Currie, M., Gerkin, R.~C., Gingell, S., Gleeson, P., Gordon, R., Hasani, R.~M., Idili, G., Khayrulin, S., Lung, D., Palyanov, A., Watts, M., and Larson, S.~D. (2018).
\newblock Openworm: overview and recent advances in integrative biological simulation of {{\it Caenorhabditis elegans}}.
\newblock {\em Philosophical Transactions of the Royal Society B: Biological Sciences}, 373(1758):20170382.

\bibitem[Segalin et~al., 2021]{Segalin2021}
Segalin, C., Williams, J., Karigo, T., Hui, M., Zelikowsky, M., Sun, J.~J., Perona, P., Anderson, D.~J., and Kennedy, A. (2021).
\newblock The mouse action recognition system (mars) software pipeline for automated analysis of social behaviors in mice.
\newblock {\em eLife}, 10:e63720.

\bibitem[Stephens et~al., 2008]{stephens2008}
Stephens, G.~J., Johnson-Kerner, B., Bialek, W., and Ryu, W.~S. (2008).
\newblock Dimensionality and dynamics in the behavior of c. elegans.
\newblock {\em PLOS Computational Biology}, 4(4):1--10.

\bibitem[Szigeti et~al., 2014]{openworm2014}
Szigeti, B., Gleeson, P., Vella, M., Khayrulin, S., Palyanov, A., Hokanson, J., Currie, M., Cantarelli, M., Idili, G., and Larson, S. (2014).
\newblock Openworm: an open-science approach to modeling {{\it Caenorhabditis elegans}}.
\newblock {\em Frontiers in Computational Neuroscience}, 8.

\bibitem[Tillmann et~al., 2024]{Tillmann2024}
Tillmann, J.~F., Hsu, A.~I., Schwarz, M.~K., and Yttri, E.~A. (2024).
\newblock A-soid, an active-learning platform for expert-guided, data-efficient discovery of behavior.
\newblock {\em Nature Methods}, 21(4):703--711.

\bibitem[White et~al., 1986]{white1986structure}
White, J.~G., Southgate, E., Thomson, J.~N., Brenner, S., et~al. (1986).
\newblock The structure of the nervous system of the nematode {{\it Caenorhabditis elegans}}.
\newblock {\em Philos Trans R Soc Lond B Biol Sci}, 314(1165):1--340.

\end{thebibliography}

\end{document}